\documentclass[prd,reprint,superscriptaddress]{revtex4-1}
\usepackage{amsmath}
\usepackage{amssymb}
\usepackage{graphicx}
\usepackage[caption=false]{subfig}
\usepackage{enumitem}
\usepackage{color}
\usepackage[percent]{overpic}
\usepackage{bm}
\usepackage{rotating}
\usepackage{hyperref}
\usepackage{pbox}
\usepackage{array}
\usepackage{physics}
\usepackage{mathtools}
\usepackage{enumerate}
\usepackage{leftidx}
\usepackage[normalem]{ulem}

\newcommand{\ii}{\mathrm{i}}

\newcommand{\tj}[6]{ \begin{pmatrix}
		#1 & #2 & #3 \\
		#4 & #5 & #6 
\end{pmatrix}}

\usepackage{suffix}
\DeclarePairedDelimiterX\MeijerM[3]{\lparen}{\rparen}%
{\begin{smallmatrix}#1 \\ #2\end{smallmatrix}\delimsize\vert\,#3}

\newcommand\MeijerG[8][]{%
	G^{\,#2,#3}_{#4,#5}\MeijerM[#1]{#6}{#7}{#8}}

\WithSuffix\newcommand\MeijerG*[7]{%
	G^{\,#1,#2}_{#3,#4}\MeijerM*{#5}{#6}{#7}}

\allowdisplaybreaks

\usepackage[export]{adjustbox}

\begin{document}
	
	\title{Light, matter, and quantum randomness generation:\\ A relativistic quantum information perspective}

	\author{Richard Lopp}
	\affiliation{Department of Applied Mathematics, University of Waterloo, Waterloo, Ontario, N2L 3G1, Canada}
	\affiliation{Institute for Quantum Computing, University of Waterloo, Waterloo, Ontario, N2L 3G1, Canada}
	\author{Eduardo Mart\'{i}n-Mart\'{i}nez}
	\affiliation{Department of Applied Mathematics, University of Waterloo, Waterloo, Ontario, N2L 3G1, Canada}
	\affiliation{Institute for Quantum Computing, University of Waterloo, Waterloo, Ontario, N2L 3G1, Canada}
	\affiliation{Perimeter Institute for Theoretical Physics, 31 Caroline St N, Waterloo, Ontario, N2L 2Y5, Canada}
	
	
	\begin{abstract}
		

		
		We study how quantum randomness generation based on unbiased measurements on a hydrogen-like
atom can get compromised by the unavoidable coupling of the atom with the electromagnetic field.  We improve on previous literature by analyzing the light-atom interaction in 3+1 dimensions with no single-mode or rotating-wave approximations and taking into account the non-pointlike nature of the atom, its orbital structure, and the exchanges of angular momentum between atom and field.  We show that preparing the atom in the ground state in the presence of no field excitations is not universally the safest state to generate randomness.

	\end{abstract}
	
	\maketitle
	
	\section{Introduction}
	Randomness is in itself a valuable resource for vastly different fields of science spanning game theory, chaos theory and cryptography. However, classical sources cannot generate true randomness since they might depend on prior information \cite{Frauchiger2013}. After all, classical mechanics is a deterministic theory and thus predictable. On the other hand, quantum theory provides a fundamental source of randomness. For example, the outcome of an unbiased measurement of an observable of a quantum system in a basis complementary to the basis in which it was prepared is  \textit{a priori} unpredictable.
	
	In that sense, one can think of a plethora of quantum systems that one could use for extracting randomness. The majority of current quantum random number generators is of optical nature, e.g. photon counting or phase noise of lasers. Another major branch consists of electronic setups, e.g. noise generation in Zener diodes or electronic shot noise \cite{Collantes2017}.
	Both branches share the feature that the system used to generate quantum randomness is fundamentally and intrinsically coupled to the electromagnetic field. As such, the quantum system can become correlated with the electromagnetic field. In principle, those correlations can be exploited by adversaries to remotely make an educated guess on the outcome of the measurement without having physical access to the quantum system. In order to understand how randomness extraction can get compromised by the coupling between the quantum system used for randomness generation and the electromagnetic field, we want to study a very simple example from atomic physics: preparing a hydrogen-like atom in a given state and measuring in a complementary basis.  
	

	
	One may think that if the preparation and the measurement are done fast enough, no information about the outcome of the measurement can possibly be leaked to the electromagnetic field; given that the atom is in its ground state (so as to minimize spontaneous emission) and placed in the vacuum of the electromagnetic field in absence of charges or currents.
	In fact, this first intuition happens to be confirmed under the common approximations in Quantum Optics (QO), namely the rotating-wave approximation (RWA) and single-mode approximation (SMA). A quick quantum-optical calculation shows that an adversary cannot increase their chances of guessing the outcome of the measurement correctly if the joint state of atom and field is in its ground state. 
	
	However, this intuition, and the calculation that backs it up, are not revealing the full story: If the coupling strength between the atom and the field is strong enough (strong-coupling in QO \cite{2004PhRvA69f2320B, 2010ApPhL96i3502C,2004Natur431162W} or ultra-strong coupling in superconducting circuits \cite{ultra}), or if the time between preparation and measurement is short enough, the most common approximations in QO (that happen to violate the local covariance of the interaction and thus render QO non-relativistic \cite{Martin-Martinez2015Causality}), such as the single mode approximation, break down. 

	An idealized model on a fully relativistic footing has been analyzed already for the case of a massless scalar field in 1+1 dimensions \cite{Thinh2016}. There, a simplified atom is modeled as an Unruh-DeWitt detector \cite{PhysRevD.14.870, deWitt}. 
	It was found that information is always leaked to the quantum field due to interactions of the atom-field system which entangle the atom and the field even when they start in their respective ground states, and even if the time between preparation and measurement is small. Moreover it turned out that a superposition of ground and excited state is the optimal state in maximizing the randomness extracted.
	
	It is important to clarify how our work is distinguished from previous work on the same issue \cite{Thinh2016}. Namely, in this study we go beyond a simple scalar 1+1 dimensional field-atom toy model employed in past literature, and we consider a fully-featured hydrogen-like atom interacting with a quantum electromagnetic field via a dipole coupling in 3+1 dimensions, with the appropriate physical values for all the fundamental constants in the problem. This allows us to capture in the model the inherently anisotropic nature of the atomic transitions, as well as the exchange of angular momentum between the atom and the field, both argued as weak points of previous literature studies.  We also consider a wider variety of time dependence of the interaction strength with time to model adiabatic versus sudden effects. These are the main novel points of this work. 
	
	
	The structure is as follows: In Section \ref{sec:general} we will provide the physical background and the setup. We will also give a measure of randomness.
	The results for the amount of randomness generated are presented in Section \ref{results} for the different considered time-dependent couplings between atom and quantum field as well as the comparison to the scalar field models of previous studies. Finally, in Section \ref{conclusion} we will summarize and discuss the results.
	
	\section{Setup}\label{sec:general}
	In the following, we will use natural units ($c=\hbar=1$) and the Minkowski metric $\eta=\text{diag}(-,+,+,+)$.
	We will consider a fully featured hydrogen-like atom coupled dipolarly to the electromagnetic field. We will not choose any of the usual approximations. Concretely we will not use the rotating-wave approximation \cite{QObook} (whose limitations were pointed out in \cite{Martin-Martinez2015Causality}) and we will not consider the atom to be point-like, and instead take into account its fully-featured orbital structure.
	We will focus on the randomness that can be extracted from a general electric dipolar transition between two levels of the atom: a ground state $\ket g_{\!\textsc{a}}$ and an excited state $\ket e_{\!\textsc{a}}$.
	
	For the state of the field, an intuitive approach would be to consider the case where no field excitations are present near the atom if we wanted the atom  to not be correlated with the field. If there were `field quanta' around the atom, surely the probability of finding the atom in one or another state would be biased towards the excited state (through photon absorption with the field) and an adversary could use that to predict the outcome of a measurement on the atom with more than 50\% accuracy, thus compromising the extraction of randomness from the atom. We would expect then that preparing the field in the vacuum state would be the best way to circumvent the bias of the probability to find the atom in the ground or excited state. For these reasons, in the same spirit as in \cite{Thinh2016}, we will consider that the electromagnetic field is in the vacuum state. However, even in the vacuum we expect atom-field interaction to create correlations between atom and field, which in turn can reduce the extracted randomness.

	Previous studies of randomness generation in atomic systems coupled to quantum fields used a simplified model of light-matter interaction (the Unruh-DeWitt model) and generic spherically symmetric smearing functions. These approaches did not account for anisotropies in the spatial distribution of the atomic wave functions of the orbitals considered for the transition, and, furthermore, they did not take into account effects coming from the exchange of orbital angular momentum between the quantum field and the atom. We will include these aspects by considering the full features of the atom.

	The dipole coupling as leading multipole term between an atom and the electromagnetic field is given by the following interaction Hamiltonian
	\begin{equation}
	\hat H_I=e\hat{\bm  x}\cdot \hat{\bm E}(\bm   x,t).
	\end{equation}
	where $\hat{\bm x}$ is the position operator of the electron with charge $e$ in the hydrogen-like atom, and $\hat{\bm E}$ is the electric field. If we express this interaction Hamiltonian in the position representation (described in full detail in section II of \cite{Pozas2016}), we obtain that the Hamiltonian, in the interaction picture, can be written as 
	\begin{align}
	\hat H_I(t)&=e ~\leftidx{_\textsc{a}\!\!}{\bra e \hat{\bm  x}\cdot \hat{\bm E}(\bm   x,t) \ket g}{_\textsc{\!a}} e^{\ii \Omega t} \ket e_{\!\textsc{a}}\!\!\bra g + \text{H.c.}\nonumber\\
	&= \int_{\mathbb{R}^3} \dd \bm x \left( \bm F(\bm x) \cdot \hat{\bm E}(\bm x, t) e^{\ii \Omega t} \ket e_{\!\textsc{a}}\!\! \bra g + \text{H.c.} \right) \nonumber\\
	&= \int_{\mathbb{R}^3} \dd^3\bm{x} ~\hat{\bm d}(\bm x , t) \cdot \hat{\bm E}(\bm x, t).
	\end{align}
	Here, $\Omega$ is the energy gap between the energy eigenstates corresponding to the two orbitals considered, and the second equality holds by insertion of unity resolved in the position eigenbasis. The dipole moment $\hat{\bm d}$ (restricted to the two relevant orbitals between which it mediates the particular transition we study) takes the form
	\begin{equation}
	\hat{\bm d}(\bm x, t)=e \left( \bm F(\bm x) e^{\ii \Omega t} \hat{\sigma}^+ +\bm F^*(\bm x) e^{- \ii \Omega t} \hat{\sigma}^- \right),
	\end{equation}
	where $\hat{\sigma}^+=\ket{e}_{\!\textsc{a}}\!\!\bra{g}$, $\hat{\sigma}^-=\ket{g}_{\!\textsc{a}}\!\!\bra{e}$ are  SU(2) ladder operators. Finally, the spatial smearing vector $\bm F(\bm x)$ is fixed by the wave functions of the orbitals involved in the atomic transition. Explicitly, as shown in \cite{Pozas2016}, the smearing vector takes the form 
	\begin{equation}
	\bm F(\bm x)=\Psi_e^*(\bm x) \bm x \Psi_g(\bm x),
	\end{equation} 
	where $\Psi_g(\bm x)$ and $\Psi_e(\bm x)$ are, respectively, the orbital wave functions of the ground and excited states of the atomic transition considered.
	To study interactions at finite times, we will add a switching function $\chi(t)$ controlling the coupling strength as a function of time such that
	\begin{align}
	H_I (t)= \chi(t) \int_{\mathbb{R}^3} \dd^3\bm{x} ~\hat{\bm d}(\bm x , t) \cdot \hat{\bm E}(\bm x, t).
	\end{align}
	
	From a practical perspective, the switching function enables us to let the boundaries of the integration of the time evolution to $\pm \infty$. We also assume that in the asymptotic past and future atom and field are uncoupled, i.e. the switching function falls off rapidly enough or has compact support. Moreover, from a physical side it can be thought of as a way to account for the finite time between preparation and measurement: a compactly supported switching function sets a clear time stamp of the preparation time (the initial interaction time after preparation) and the measurement time (the amount of time from preparation to measurement). In addition, a switching function allows us to model more accurately experimental setups. For instance, we could initially place the atom inside a small enough cavity such that the lowest energy mode of the cavity determined by the IR cut-off lies well above the energy gap of the atom. In that case, the atom being placed inside a Faraday cage effectively does not interact with the cavity field nor with the field outside. When we subsequently remove the cavity, we create a coupling between atom and field. The interaction time is finite if we  bring back the cavity. This could correspond to a sudden top-hat switching function given the cavity is removed and brought back quick enough. Another example modeled by a Gaussian switching function could correspond to an atom moving transversely through a cavity since the ground transversal mode of a cavity has a Gaussian-shaped amplitude profile. More generally, even though we study the behavior of atom-field interaction in free space in the following, one can model the evolution in a highly controlled light-matter interaction setup. It is possible to temporally vary the coupling strength between a superconducting qubit and the electromagnetic field inside a microwave cavity. In that way one can design a range of switching functions \cite{SCqubit2010,Thinh2016}.

	For our purposes, we expand the electric field operator into plane-wave modes of momentum $\bm k$ and polarization $s$, with their respective creation and annihilation operators $\hat{a}^{\dagger}_{\bm k, s}$ and $\hat{a}_{\bm k, s}$, satisfying the canonical equal time commutation relations. In this form, the field operator can be written as
	\begin{equation}
	\hat{\bm E}(\bm x, t)\!=\! \sum_{i=1}^2\! \int \!\frac{\text{d}^3\bm{k}}{(2 \pi)^{3/2}} \sqrt{\frac{|\bm k|}{2}} \left( -\ii \hat{a}_{\bm k, s_i} \bm \epsilon(\bm k, s_i) e^{\ii  \mathsf{k} \cdot  \mathsf{x}} + \text{H.c.} \right),
	\end{equation}
	where $\mathsf{k}$ and $\mathsf{x}$ are $4$-vectors and we denoted as $\bm \epsilon(\bm k, s_1)$ and $\bm \epsilon(\bm k, s_2)$ an arbitrary set of two independent transverse polarization vectors ($\bm k\cdot\bm \epsilon(\bm k, s_i) =0$).

	
	The time evolution of the coupled system of atom and quantum field is captured by the unitary operator $\hat U$ acting on the initial joint state of the system $\hat \rho_{i}$ such that after the interaction the joint system is in the state
	\begin{align}
	\hat \rho_{\textsc{af}} &=\tilde{\ket{\Psi}}_{\!\textsc{af}}\! \tilde{\bra{\Psi}}=\hat U \hat \rho_i \hat{U}^{\dagger}, 
	\end{align}
	where
	\begin{equation}
	\hat{U} =\mathcal{T} \exp( -\ii \int_{-\infty}^{\infty} \text{d}t \hat{H}_I(t)).\label{evolution}
	\end{equation}
	$\mathcal{T}$ denotes the time-ordering operation. We will assume that initially field and detector are uncorrelated and hence in a product state of the form
	\begin{align}
	\hat \rho_{ i}=\ket \Psi_\textsc{\!a}\!\! \bra \Psi \otimes \ket 0_\textsc{\!f}\!\! \bra 0,
	\end{align}
	where $\ket \Psi_\textsc{\!a}$ is some arbitrary superposition of the energy eigenstates of the atom, and, as noted before, the field is in the vacuum state.
	
	Generating randomness from an atomic probe (i.e., two energy levels) is conceptually easy: one prepares an initial state of the atom, and then performs a  von-Neumann measurement in a complementary basis. However, even theoretically, this protocol for extracting randomness was too na{i}ve: atoms are always intrinsically coupled to the electromagnetic field. In between the preparation of the atom and the projective measurement, the atom interacts with the electromagnetic field which will generally correlate both, giving an adversary with access to the field means to make an educated guess on the result of the measurement. Contrary to intuition, the acquisition of correlations between the field and the state of the atom can happen even if the time between preparation and projection is small, and even if both atom and field start in the ground state \cite{Thinh2016, Martin-Martinez2015Causality}. These correlations serve as a bias which can be exploited by an adversary who has access to the field to infer the measurement outcome better than just by chance. In order to prevent this, two options are at disposal. First, one can try to change the initial state of the atom to minimize these correlations, and secondly a different measurement basis might allow to re-establish an unbiased situation.

	Let us formalize the problem: The joint system (atom-field) is prepared in its initial state in some basis at some time. Following preparation, atom and field interact with each other, and after some time $T$ the von-Neumann measurement $\{\hat P_x\}$ will be performed in some other arbitrary basis on the atom with the objective of generating randomness. From this measurement one obtains the result $x=\{ 0,1 \}$, eigenvalues of some observable $\hat X$. This yields the new total state \mbox{$\hat \rho_{\hat{\textsc{x}} \textsc{f}}^x=\ket x_\textsc{\!a}\!\! \bra x \otimes \hat\tau_{\textsc{f}}^x$}. The state of the field after the projection ($\hat\tau_{\textsc{f}}^x$) can be obtained by tracing out the atom and it is given by
	\begin{align}
	\hat\tau_{\textsc{f}}^x=\frac{\tr_{\textsc{a}}\!\big(\hat P_x \hat \rho_{\textsc{af}}\big)}{\tr\big(\hat P_x \hat \rho_{\textsc{af}}\big)}.
	\end{align}
	This state can possibly be accessed by an adversary in order to infer the measurement result $x$.
	
	The conditional min-entropy \cite{Reyzin,Thinh2016, Koenig2008} will be used to quantify a lower bound on the extracted randomness by an adversary with access to the quantum field after the initial measurement and is defined as
	\begin{equation}
	H_{\text{min}}(X | F)_{\hat{\rho}_{\textsc{x} \textsc{f}}}=-\log_2 \left[P_g(X | F)_{\hat{\rho}_{\textsc{x} \textsc{f}}}\right],
	\end{equation}
	where $P_g( X | F)_{\hat \rho_{\textsc{xf}}}$ denotes the probability of guessing correctly the outcome of a measurement on the random variable $X$ associated to the observable $\hat X$ given access to the partial state of the field $F$, and where \mbox{$\hat \rho_{{\textsc{x}} \textsc{f}}= \sum_x p_{{X}}(x) \hat \rho^x_{{\textsc{x}} \textsc{f}}$} is the statistical ensemble of the possible measurement outcomes.
	
	The choice of the min-entropy as a figure of merit to quantify randomness is justified by the following rationale: 
	Since the min-entropy takes the value $k$ if all outcomes of a distribution occur  at most with probability $2^{-k}$, we have a necessary condition to generate $k$ random bits from the distribution. More generally, the distribution only has to be $\epsilon$-close to a distribution that has min-entropy $k$ \cite{Random2}.
	
	The min-entropy also constitutes a much better estimator of randomness than the Shannon entropy, which coincides with the min-entropy for homogeneous (flat) distributions. The reason is that the Shannon entropy yields the gain of information about a distribution obtained per individual sampling after taking the average over (asymptotically infinitely) many independent samples, whereas the min-entropy quantifies the gain of information when taking only one sample in the `worst-case' scenario \cite{Randomextractor}. Due to this averaging, we cannot conclude that having access to a random variable with a high Shannon entropy  we are in possession of a good randomness source. Therefore, the min-entropy functions as a more conservative estimator of randomness. Indeed, the min-entropy is always bounded from above by the Shannon entropy. Accordingly, it is known that often the Shannon entropy significantly overestimates the amount of randomness obtainable from a random variable \cite{Random2}.
	
	Another point to take into account is the fact that the quantum field is infinite-dimensional. From the point of view of randomness extraction, the issue of the infinite-dimensionality of the field can be reduced to a problem of finite number of degrees of freedom since, by construction, the atom possesses a finite number of energy eigenstates.  For example, in this paper we will consider the conservative case where we quantify the randomness that can be extracted from only two levels connected by an electric dipole transition, such that the field can excite the ground state of the atom only to one higher energy state (in the same fashion as it was done in \cite{Thinh2016} for a scalar field). We can write the final pure state of the joint system after interaction via Schmidt decomposition as
	\begin{align}
	\tilde{\ket{\Psi}}_{\!\textsc{af}}=\sqrt{\lambda_0}\ket 0_\textsc{\!a} \otimes \ket{f_0}_\textsc{\!f} + \sqrt{\lambda_1}\ket 1_\textsc{\!a} \otimes \ket{f_1}_\textsc{\!f},
	\end{align}
	where $\{\ket{i}_\textsc{\!a}\}$ are the eigenstates of the observable $\hat X$ (not necessarily the energy eigenstates of the atom), and $\{\ket{f_i}_\textsc{\!f} \}$ are two orthonormal basis states out of the field's infinite-dimensional Hilbert space. \textit{A priori} these basis states of the field are not known, and their precise form is not even needed to arrive at an analytic expression for the amount of generated randomness. If the adversary wants to implement a protocol to optimize the guessing probability for the measurement outcome, then they would indeed need to construct $\{\ket{f_i}_\textsc{\!f} \}$ by a Schmidt decomposition algorithm, and may involve many (possibly infinite) field modes. However, we do not concern ourselves with finding that specific decomposition as doing so is the adversary's task. Rather, our objective is to reduce their ability to make educated guesses on the randomly generated data by probing the field. Thus, we should keep the most conservative assumptions on the adversary's ability. Considering that the min-entropy is invariant under local isometries \cite{Renner2010}, we can devise a unitary operation that transfers the information from the field to an ancillary qubit E in possession of the adversary, e.g. swapping entanglement between field and E. Therefore, the new final joint state reads
	\begin{align}
	\tilde{\ket{\Psi}}_{\!\textsc{ae}}=\sqrt{\lambda_0}\ket 0_\textsc{\!a} \otimes \ket{0}_\textsc{\!e} + \sqrt{\lambda_1}\ket 1_\textsc{\!a} \otimes \ket{1}_\textsc{\!e}.
	\end{align}
	Accordingly, after the von-Neumann measurement on atom A, the ensemble corresponding to the different outcomes is
	\begin{align}
	\hat \rho_{\textsc{ae}}^x=\sum_x \ket{x}_\textsc{\!a}\!\!\bra{x} \otimes \hat\tau_\textsc{e}^x,
	\end{align}
	with the qubit E being, for the outcome $x$, in the state 
	\begin{align}
	\hat\tau_\textsc{e}^x=\frac{\tr_\textsc{a}\!\big(\hat P_x \hat \rho_{\textsc{ae}}\big)}{\tr\big(\hat P_x \hat \rho_{\textsc{ae}}\big)}.
	\end{align}
	The probability of guessing correctly the outcome is equivalent to the optimal success probability of the adversary to distinguish the states of the qubit  $\hat\tau_\textsc{e}^x$:
	\begin{align}
	P_g\big( X|E\big)_{\hat \rho_{\textsc{ae}}}&=\max_{\hat{\mathcal{E}}} \sum_x p_{ X}(x)\, \leftidx{_\textsc{e}\!\!}{\bra{x}} \hat{\mathcal{E}}(\hat\tau_\textsc{e}^x)\ket{x}_\textsc{\!e}\nonumber \\
	&=\max_{\hat{\Pi}_x}\sum_x p_{ X}(x) \tr(\hat{\Pi}_x \hat\tau_\textsc{e}^x),
	\end{align}
	where we optimize over CPTP maps $\hat{\mathcal{E}}$ or equivalently over POVMs $\{\hat{\Pi}_x=\hat{\mathcal{E}^\dagger}(\ket x_\textsc{\!e}\!\! \bra x) \}$. It is assumed that the adversary knows the measurement basis given by $\hat X$, rendering the adversary more powerful. By the Helstrom bound \cite{Helstrom} for the minimum-error probability of distinguishing two states by optimizing over POVMs we find
	\begin{align}
	P_g\big( X|E\big)_{\hat \rho_{\textsc{ae}}}&=\frac{1}{2}\left(1+ \lVert p_{ X}(0)\hat\tau_\textsc{e}^0 - p_{ X}(1) \hat\tau_\textsc{e}^1  \rVert_1 \right), \label{Pg}
	\end{align}
	where $\lVert \hat O \rVert_1 = \tr \sqrt{\hat{O}^{\dagger} \hat O}$ is the Schatten $1$-norm. Counteracting the adversary to yield the maximum (denoted by superscript $*$) amount of randomness $H_{\text{min}}^*$ which can be extracted from the atom, we have to optimize over all von-Neumann measurements on atom A. Any arbitrary complex two-dimensional projector decomposition can be written as a linear combination of projectors of the form $\hat P_i = \ket{m_i}_\textsc{\!a}\!\! \bra{m_i}$ with
	\begin{align}
	\ket{m_0}_\textsc{\!a}=\cos \theta \ket 0_\textsc{\!a} + e^{\ii \phi} \sin \theta \ket 1_\textsc{\!a}, \nonumber \\
	\ket{m_1}_\textsc{\!a}=\sin \theta \ket 0_\textsc{\!a} - e^{\ii \phi} \cos \theta \ket 1_\textsc{\!a}.
	\end{align}
	Then we find that $p_X(x) \hat \tau_\textsc{e}^x=\ket{n_x}_\textsc{\!e}\!\!\bra{n_x}$, where
	\begin{align}
	\ket{n_0}_\textsc{\!e}=\sqrt{\lambda_0}\, \leftidx{_\textsc{a}\!\!}{\braket{m_0}{ 0}}{_\textsc{\!a}} \ket{0}_\textsc{\!e} + \sqrt{\lambda_1}\, \leftidx{_\textsc{a}\!\!}{\braket{m_0}{ 1}}{_\textsc{\!a}} \ket{1}_\textsc{\!e} , \nonumber \\
	\ket{n_1}_\textsc{\!e}=\sqrt{\lambda_0}\, \leftidx{_\textsc{a}\!\!}{\braket{m_1}{ 0}}_\textsc{\!a} \ket{0}_\textsc{\!e} + \sqrt{\lambda_1}\,\leftidx{_\textsc{a}\!\!}{\braket{m_1}{ 1}}{_\textsc{\!a}} \ket{1}_\textsc{\!e}.
	\end{align}
	This allows us to write the optimized guessing probability, by using Eq.\! \eqref{Pg}, as
	\begin{align}
	P^*_g\big( X|E\big)_{\hat \rho_{\textsc{ae}}}&=\min_{\{\ket{m_i}_\textsc{\!a}\}} \frac{1}{2}\left(1+ \sqrt{1-4 \left|\leftidx{_\textsc{e}\!\!}{\braket{n_0}{n_1}}{_\textsc{\!e}} \right|^2} \right)\nonumber \\
	&= \frac{1}{2}\left(1+ \sqrt{1-(\lambda_0 - \lambda_1)^2} \right)\nonumber \\
	&=\frac{1}{2}+ \sqrt{\frac{1}{4}-\left(\frac{1}{2}\tr\left(\hat \rho_\textsc{a}^2\right)-\frac{1}{4}\right)} ,
	\end{align}
	where  $\hat \rho_\textsc{a}$ is the reduced density matrix of the atom after its interaction with the field from preparation to measurement.
	Finally we find the expression for the optimized min-entropy:
	\begin{equation}
	H^*_{\text{min}}=-\log_2\left(\frac{1}{2}+\sqrt{\frac{1-\tr\left(\hat \rho_\textsc{a}^2\right)}{2}} \right). \label{hmin}
	\end{equation}
	Thus, it is sufficient to know the state of the atom after the interaction to fully quantify the extractable randomness. This measure of randomness generation can also be viewed from the point of view of device-independent random number generation and quantum key distribution -- see, for instance, \cite{Colbeck, Piro, Lunghi, Vidick}. In that context the min-entropy gives an quantitative estimate allowing to certify whether the output is truly random whilst treating the random number generator as a black box. 
	
	\section{Results}\label{results}
	\subsection{Final atomic state}
	
	The time-evolved state will be calculated by a perturbative Dyson expansion of \eqref{evolution}, granted the relevant parameters are small enough:
	\begin{equation}
	\hat U=\openone\underbrace{-\ii\int_{-\infty}^{\infty}\!\!\! \dd t\,\hat H_I(t)}_{\hat U^{(1)}}\underbrace{-\!\!\int_{-\infty}^{\infty}\!\!\!\!\text{d}t\int_{-\infty}^{t}\!\!\!\!\!\!\text{d}t^{\prime}\,\hat H_I(t)\hat H_I\left(t^{\prime}\right)}_{\hat U^{(2)}}+\dots\label{eq:evo}
	\end{equation}
	Thus, to second order in the coupling constant $e$ the evolved state takes the form
	\begin{align}
	\hat \rho_\textsc{af}&= \hat \rho_i + \hat U^{(1)} \hat \rho_i + \hat \rho_i \hat U^{(1) \dagger} \nonumber\\
	&\quad +  \hat U^{(2)} \hat \rho_i + \hat \rho_i \hat U^{(2) \dagger} +  \hat U^{(1)} \hat \rho_i \hat U^{(1) \dagger} + \mathcal{O}(e^3).
	\end{align}
	We assume that the initial state of the field is its ground state $\ket 0_{\!\textsc{f}}$ and the atom is in some superposition of its energy eigenstates $\ket \Psi_{\!\textsc{a}} = a \ket g_{\!\textsc{a}} + \sqrt{1-a^2} \ket e_{\!\textsc{a}}$, where we restrict $a$ to be real and $a=1$ ($a=0$) corresponds to the ground state (excited state). Hence the initial state of the atom reads 
	\begin{align}
	\hat \rho_{\textsc{a},i}= \begin{pmatrix}
	a^2 & a \sqrt{1-a^2} \\
	a \sqrt{1-a^2} & 1-a^2 \\
	\end{pmatrix} \label{initi}
	\end{align}
	in the $\{\ket g_{\!\textsc{a}}, \ket e_{\!\textsc{a}}\}$ basis. Then after interaction between atom and field, and before subsequent measurement on the atom, the final atomic state reads to second order
	\begin{equation}
	\hat \rho_\textsc{a}=\hat \rho_{\textsc{a},i}  + \underbrace{\tr_{\textsc{f}}\left( \hat U^{(1)} \hat \rho_i \hat U^{(1) \dagger}\right) + \left(\tr_{\textsc{f}}\left( \hat U^{(2)} \hat \rho_i\right) + \text{H.c.}\right)}_{\Delta \hat \rho}. \label{atomic rho}
	\end{equation}
	Therefore we call $\Delta \hat \rho$ the correction to the initial state carrying the time evolution of the atom to leading order. Note that in \eqref{atomic rho} there are no first order terms. This is because for the vacuum state $\tr_{\textsc{f}}\left( \hat U^{(1)} \hat \rho_i\right)=0$.

	In Appendix~\ref{app1} the derivation is explicitly shown in general form for arbitrary atomic transitions and switching functions. In particular, the final results for the exemplary $1s\rightarrow 2p_z$ atomic transition are given for the following switching functions (see Appendix~\ref{partic}): 1)  Gaussian switching  $\chi^\textsc{g}(t)=e^{-t^2/\sigma^2}$, 2) sudden Heaviside top-hat switching $ \chi^\textsc{s}(t)=\Theta(t)\Theta(-t+\sigma)$ and 3) Dirac delta switching $\chi^\textsc{d}(t)=C \delta(t)$, where $\sigma$ is the interaction time scale and the constant $C$ is needed for correct dimensionality. This yields for the change in the atomic state to second order in perturbation theory respectively
	\begin{widetext}
		\begin{align}
		\Delta \hat \rho^\textsc{g}=&\frac{ 24576 (a_0 e \sigma)^2}{ \pi}\int_0^{\infty} \dd |\bm k| \frac{|\bm k|^3 e^{-\frac{1}{2} \sigma^2 (|\bm k|+ \Omega)^2} }{\left(4 a_0^2 |\bm k|^2 +9\right)^6} \left\{ 2 \left[(1-a^2) e^{2 |\bm k| \sigma^2 \Omega}-a^2 \right] \begin{pmatrix}
		1&0 \\
		0 & -1  \\
		\end{pmatrix} \right. \nonumber \\
		&\quad+\left. \left( a \sqrt{1-a^2} \left[e^{2 |\bm k| \sigma^2 \Omega} \text{erf}\left(\frac{\ii \sigma (|\bm k| - \Omega)}{\sqrt{2}} \right)- \text{erf}\left(\frac{\ii \sigma (|\bm k| + \Omega)}{\sqrt{2}} \right) -\left(1- e^{|\bm k| \sigma^2 \Omega}\right)^2 \right] \begin{pmatrix}
		0&0 \\
		1 & 0  \\
		\end{pmatrix}
		+ \text{H.c.} \right) \right\},\label{gaussian} \\
		\Delta \hat \rho^\textsc{s}=& \frac{49152 (a_0 e)^2}{\pi^2} \int_0^{\infty} \!\dd |\bm k| \frac{ |\bm k|^3 }{\left( 4 a_0^2 |\bm k|^2 +9 \right)^6 (|\bm k|^2-\Omega^2)^2} \left\{\vphantom{\begin{pmatrix}
			1 & 0 \\
			0 & -1 \\
			\end{pmatrix}} \left[2a^2 (|\bm k|-\Omega )^2 \cos  (\sigma  (|\bm k|+\Omega )) \right.\right. \nonumber \\
		&\quad\quad \left.\left. +2\left(1-2 a^2\right) \left(|\bm k|^2+\Omega^2\right)+4 |\bm k| \Omega+2(a^2-1\right)  \cos  (\sigma  (|\bm k|-\Omega ))(|\bm k|+\Omega )^2 \right]  \begin{pmatrix}
		1 & 0 \\
		0 & -1 \\
		\end{pmatrix}  \nonumber \\
		&\,\quad+ \left( \vphantom{\begin{pmatrix}
			1 & 0 \\
			0 & -1 \\
			\end{pmatrix}} \left[ e^{2 \ii \sigma \Omega}(|\bm k|^2 -\Omega^2) +|\bm k|^2 (2 \ii \sigma \Omega-1) +  4 \Omega e^{\ii \sigma \Omega} ( \Omega \cos{(|\bm k| \sigma)} - \ii |\bm k| \sin{(|\bm k| \sigma)}) - \Omega^2 (3 + 2\ii \sigma \Omega)\right] \right. \nonumber \\
		&\left. \left.\quad \quad  \times a \sqrt{1-a^2}  \begin{pmatrix}
		0 & 0 \\
		1 & 0 \\
		\end{pmatrix}  +  \text{H.c.} \right)\right\}, \label{heaviside} \\
		\Delta \hat \rho^\textsc{h}_{\Omega=0}=& \frac{512 e^2 }{295245 \pi^2}\left(1- 2 a^2 \right) \left[24-\sqrt{\pi} \MeijerG*{2}{1}{1}{3}{0}{0,5,\frac{1}{2}}{\frac{9 \sigma^2}{16 a_0^2}}\right]  \begin{pmatrix}
		1 & 0 \\
		0 & -1 \\
		\end{pmatrix} , \label{gapless} \\
		\Delta \hat \rho^\textsc{d}=& \frac{128 C^2 e^2}{10935 \pi^2 a_0^2} (1-2 a^2) \begin{pmatrix}
		1&0 \\
		0 & -1  \\
		\end{pmatrix}\label{diracfinal},
		\end{align}
	\end{widetext}
	where $\MeijerG*{m}{n}{p}{q}{a_1, \dots, a_p}{b_1, \dots, b_q}{z}$ is the Meijer G-function, erf$(z)$ is the error function and  $a_0$ is the generalized Bohr radius. Eq. \!\eqref{gapless} is obtained from \eqref{heaviside} for degenerate atomic transitions ($\Omega=0$), also called gapless sudden switching.
	
	\subsection{Extracted randomness}
	
	In the previous section we have obtained the time evolved density matrix of the atom from the time of preparation to the time when the measurement is performed for the different switching functions considered, namely Gaussian \eqref{gaussian}, sudden \eqref{heaviside}-\eqref{gapless}, and delta \eqref{diracfinal}.
	With this information at hand, we can now calculate the number of bits of randomness that can be generated with each measurement. We will present the results for Gaussian switching, gapless sudden switching and delta switching separately.
	
	The first step is to choose physically meaningful values for the parameters of the problem. As a baseline, we start with the parameters $a_0\approx2.68\cdot 10^{-4}~\text{eV}^{-1}$,  \mbox{$e\approx 137^{-1/2}\approx8.54\cdot 10^{-2}$}, $\Omega \approx 3.73~\text{eV}$. These have been chosen such that the atomic radius corresponds to the Bohr radius, $e$ to the standard electric charge in vacuum (the square root of the fine structure constant in natural units) and \mbox{$a_0 \Omega\approx 0.001$} is of the same order of a typical transition from the ground state to the first excited state in a hydrogen-like atom \cite{CODATA2014}. By varying $a_0$, $e$, $\Omega$ we will study how the generated randomness is dependent on these parameters. 
	
	\subsubsection{Gaussian switching: $\chi^\textsc{g}(t)=e^{-t^2/\sigma^2}$}
	\begin{figure*}
		\centering
		\subfloat[]{
			\centering
			\includegraphics[scale=1]{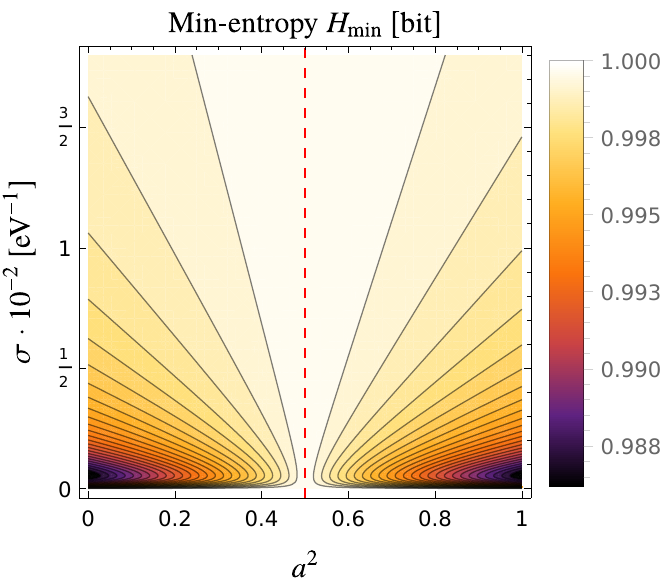}}
		\qquad \qquad \qquad 
		\subfloat[]{
			\centering
			\includegraphics[scale=1]{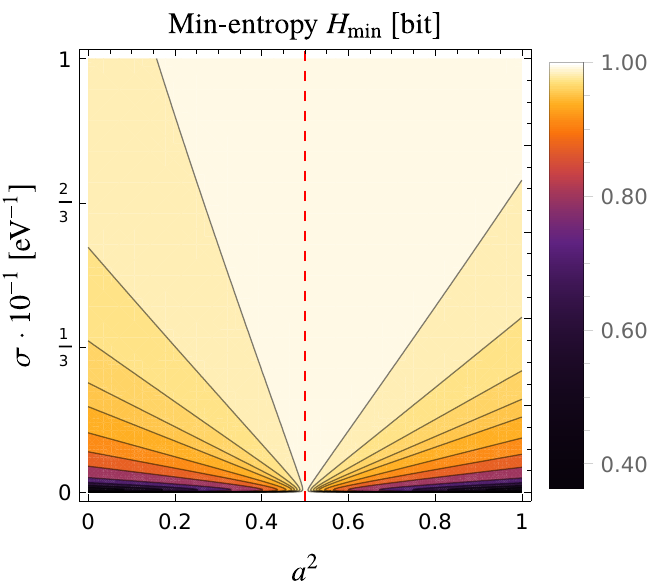}}
		\caption{Min-entropy $H_{\text{min}}$ plotted against duration of interaction $\sigma$ and $a^2$ (proportional to the $z$ component of the initial Bloch vector) for Gaussian switching with the parameters $a_0=2.68\cdot 10^{-4}~\text{eV}^{-1}$, $\Omega =3.73~\text{eV}$, and (a) free-space coupling  $ e=8.54\cdot 10^{-2}$ or (b) strong coupling $ e=5$. Recall that $a=0$ corresponds to $\ket e_{\!\textsc{a}}$ and $a=1$ to $\ket g_{\!\textsc{a}}$. $H_{\text{min}}=1~$bit coincides with maximal randomness and is in fact never absolutely reached. The highest amount of randomness can be found for an equal superposition $a= {1}/\!\sqrt{2}$ (red dashed line). The excited state and, surprisingly, the ground state are the least favored preparations for short interactions. The lack of symmetry is related to the non-homogeneous nature of the switching and it is explained in detail in Appendix~\ref{symm}. }
		\label{gauss1}
	\end{figure*}

	For a Gaussian switching function, the amount of randomness that can be generated as a function of the interaction time $\sigma$ and initial superposition parameter $a$ is shown in  Fig.~\ref{gauss1}.
	As a general feature we note that for shorter interaction times the amount of randomness is compromised more severely. In fact, we see that for the regular free-space coupling of Fig.~\ref{gauss1}a, Gaussian switching provides a good source of randomness for interaction times above $\approx 10^{-2}~\text{eV}^{-1}$, which in principle tells us that an adiabatic switching (smooth switching that depends only on one timescale, such as Gaussian) prevents the generation of atom-field correlations well enough to guarantee a reliable extraction of randomness.   However, this is not true for regimes of strong coupling: as we will comment on below, the amount of randomness extracted decays fast with the interaction strength and becomes relevant for strong coupling strengths. 
	
	Remarkably, and contrary to intuition the ground state is not the most secure choice of initial atom preparation for short interaction times. It turns out that an equal superposition of ground and excited state is most resilient and in fact yields min-entropy very close to $1$~bit. Moreover, the initial guess that the excited state of the atom may be the worst preparation (because of its probability of spontaneously decay) is not the complete picture. Surprisingly, the ground state is almost as bad a choice as the excited state in terms of generation of randomness. 
	
\begin{figure*}
		\includegraphics[scale=.7,valign=m]{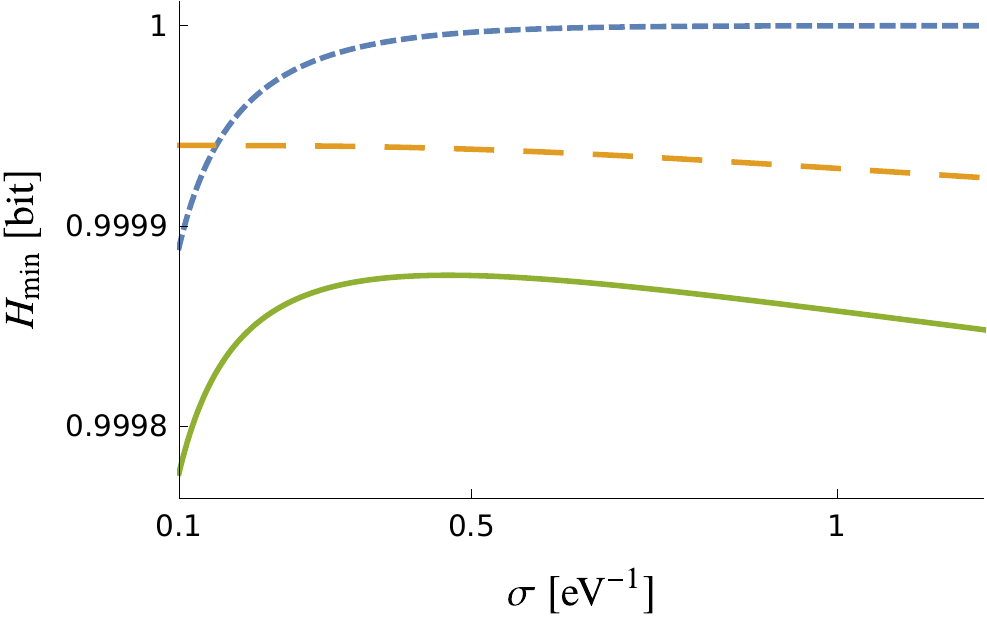}
		\includegraphics[scale=.95]{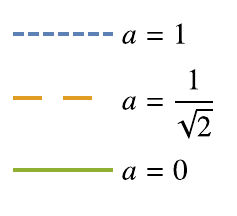}
		\caption{Min-entropy  $H_{\text{min}}$ for longer interaction times $\sigma$ with parameters  $a_0=2.68\cdot 10^{-4}~\text{eV}^{-1}$, $\Omega =3.73~\text{eV}$, $ e=8.54\cdot 10^{-2}$ for the ground ($a=1$) and excited state ($a=0$), and equal superposition ($a={1}/\!\sqrt{2}$) in the case of Gaussian switching. The ground state recovers $H_{\text{min}}=1~$bit, the other initial atomic preparations witness a fall-off of the extracted randomness. }
		\label{late time}
	\end{figure*}

	This stresses our claim: for fast randomness generation the equal superposition state provides the best possible initialization of the system. Nonetheless, as we would expect, for the late interaction time regime we recover that the ground state yields maximum randomness generation whereas all other state preparations, including excited state and equal superposition, experience a decrease in randomness for longer interaction times (see Fig.~\ref{late time}). 
	
	\begin{figure*}
		\includegraphics[scale=.95]{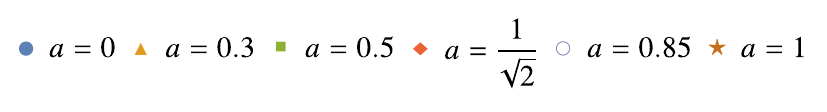}
	\hspace*{-.2cm}	\includegraphics[scale=.8,valign=l]{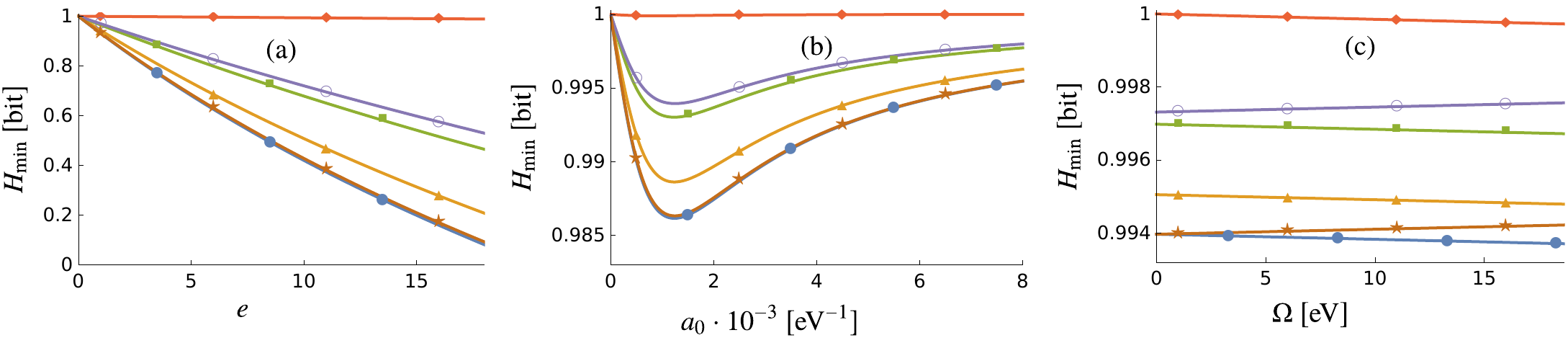}
		\caption{Extracted min-entropy as a function of (a) electric charge $e$, (b) atomic radius $a_0$, and (c) energy gap $\Omega$ for a fixed interaction time $\sigma=2.5 \cdot 10^{-3}~\text{eV}^{-1}$ and different atomic state parameters $a$ in the case of Gaussian switching. For (a) $a_0=2.68\cdot 10^{-4}~\text{eV}^{-1}$, $\Omega =3.73~\text{eV}$, in (b) $ e=8.54\cdot 10^{-2}$, $\Omega =3.73~\text{eV}$, and in (c)  $a_0=2.68\cdot 10^{-4}~\text{eV}^{-1}$, $ e=8.54\cdot 10^{-2}$.  In (a) and (b) $\sigma \Omega \approx 0.01$, and in (c) $\sigma \Omega <  0.045$, and hence all plots are beyond the validity of the RWA. For (a), as explained in Appendix~\ref{numvalues}, perturbation theory holds for $e<12.8$ (see Appendix~\ref{numvalues} to find a discussion about the whole plot domain).} 
		\label{gausspara}
	\end{figure*}
	
	In Fig.~\ref{gausspara} we show the dependence of the extracted randomness on the parameters $e$, $a_0$ and $\Omega$.  For the chosen interaction time of $\sigma=2.5 \times 10^{-3}$, we are in a regime where the rotating wave approximation is not valid, i.e. $\sigma \Omega \ll 1$.
	The stronger the coupling $e$ between atom and field the less randomness will be generated overall since it results in the enhancements of acquired atom-field correlations. The extracted randomness falls off more quickly for states which are closer to being either of the two energy eigenstates. This is particularly relevant as it is shown in Fig.~\ref{gauss1}b: in regimes of strong coupling the loss of randomness at short times can still be relatively significant for timescales of $10^{-1}~\text{eV}^{-1}$. 
	
	 One concern may be that the perturbative expansion is not valid for regimes of very low min-entropy as shown in Fig.~\ref{gauss1}b. For instance, $H_{\text{min}}=0.4~$bit means that the final atomic state has a purity of, using \eqref{hmin}, $\tr\left(\hat \rho_\textsc{a}^2\right)=0.87$. In the Appendix~\ref{numvalues} we will present numerical values for the change of the atomic state (for all switching functions), showing that perturbation theory holds for low min-entropy in our analysis up to some value of the coupling strength (depending on the switching function) for which we provide a lower bound in the appendix.

	For the dependence on the atomic radius we find that for large values of $a_0$ the generated randomness asymptotically approaches a constant value after passing through a minimum. The depth of the minimum is larger for states that are closer to either of the energy eigenstates. Hence the equal superposition of them shows to be very close to constant. 
	
	The extracted randomness decreases with larger values for the energy gap $\Omega$ for atomic states with parameter $a \leq {1}/\!\sqrt{2}$ and increases for the remaining states. Hence the ground state or in general states with the major probability of being in the ground state after preparation become more secure when the gap between the energy eigenstates increases. This is consistent with the intuition that a larger gap makes it more difficult for the ground state to get excited through a counter-rotating process (emitting excitations that could be captured by an adversary). At the same time, increasing the gap increases the probability that the excited states decayed emitting light, which in turn can be captured to infer the measurement outcome. An equal superposition state is overall most resistant to variations in these parameters and, moreover, is close to being constant in all three parameter cases.
	

	\subsubsection{Sudden switching:  $ \chi^\textsc{s}(t)=\Theta(t)\Theta(-t+\sigma)$}
	
\begin{figure*}
	\centering
	\begin{tabular}[c]{cc}
		\begin{tabular}[b]{c}
			\subfloat[  ]{\label{fig:right}%
			\hspace*{-0.4cm}	\includegraphics[scale=.9,valign=b]{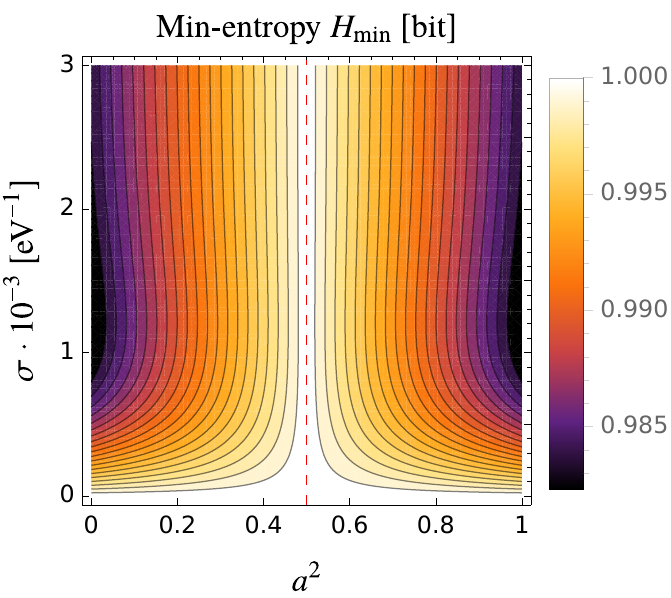}}

		\end{tabular}
	&
		\subfloat{\label{fig:left}%

			\begin{tabular}[b]{c}
					\includegraphics[scale=.95,valign=b]{leg3.pdf}\\[2mm]
			\hspace*{-.09cm}	\includegraphics[scale=.79,valign=b]{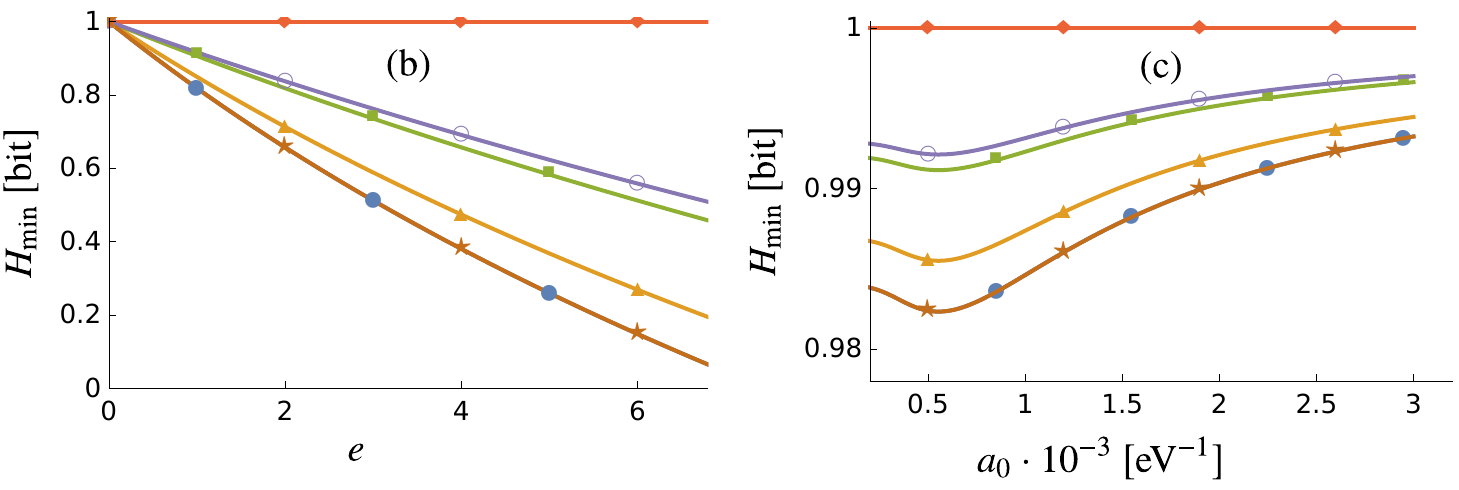}
		\end{tabular}
 }
		
	\end{tabular}
	\caption{(a) Min-entropy $H_{\text{min}}$ plotted against duration of interaction $\sigma$ and $a^2$ (proportional to the $z$ component of the Bloch vector) with the parameters $a_0=2.68\cdot 10^{-4}~\text{eV}^{-1}$, $ e=8.54\cdot 10^{-2}$ for gapless sudden switching. The red dashed line corresponds to equal superposition and yields a maximum min-entropy of $H_{\text{min}}=1~$bit. Ground and excited state witness the least amount of randomness that can be extracted. (b, c) Extracted min-entropy as a function of the parameters (b) electric charge $e$ and (c) atomic radius $a_0$ for a fixed interaction time $\sigma=2.5 \cdot 10^{-3}~\text{eV}^{-1}$ and different atomic state parameters $a$ in the case of sudden switching. In (b) we keep $a_0=2.68\cdot 10^{-4}~\text{eV}^{-1}$ fixed and for (c) we have $ e=8.54\cdot 10^{-2}$. According to the discussion in Appendix~\ref{numvalues}, in (b) perturbation theory holds safely for $e<4.7$. (see Appendix~\ref{numvalues} to find a discussion about the whole plot domain).} 
	\label{unit}
\end{figure*}

	We consider here the case of an infinitely fast switching on and off, modelled by a square function. For the sudden top-hat switching we will study the case of degenerate atomic transitions ($\Omega=0$) due to numerical simplicity. The min-entropy portraits a different picture (see Fig.~\ref{unit}a) than for the Gaussian switching function. It is still true that an equal superposition of ground and excited state is the most secure state to generate randomness for general interaction times. However, short interaction times between field and atom yield a larger min-entropy than longer interaction times. On the other hand, for later times the min-entropy varies very little with the interaction time for fixed $a$. It suggests that the amount of randomness that can be extracted takes an asymptotic value for fixed $a$. In this case, we observe that it is preferable to perform the measurement very fast in order to avoid the loss of randomness coming from the regime of long interaction times. This stands in contrast to the Gaussian switching where it is better to choose a longer interaction time between atom and electromagnetic field.
	
	From \eqref{gapless} it is obvious that to second order in perturbation theory the equal superposition provides us with a state that yields $H_{\text{min}}=1~$bit since that state is a fixed point in time evolution (does not vary in time) for the degenerate transition case. 
 We want to highlight that the symmetry around $a={1}/\!\sqrt{2}$ is exact in Fig.~\ref{unit}a. Details are given in the Appendix~\ref{symm}. There we will also comment on the exact symmetry in the case of Dirac switching.

%
%

In Fig.~\ref{unit}b and Fig.~\ref{unit}c the dependence of the min-entropy on its parameters $e$ and $a_0$ is shown for fixed times $\sigma$.
As in the case of Gaussian switching, a stronger coupling implies a decrease of extracted randomness. It also holds that states prepared close to being in an equal superposition show a slower decrease in the min-entropy than for states which are prepared close to being in an energy eigenstate.
Moreover, for small atomic radii $a_0$ the extracted randomness shows a minimum and increases then asymptotically to a constant value of the min-entropy, depending on $a$. In summary, the equal superposition provides the optimal state to extract randomness as it is in fact independent of the parameters to leading order in perturbation theory. It should be noted that because the eigenstates of the atom are degenerate ($\Omega=0$) there is no regime where  the RWA condition is satisfied.

\subsubsection{Delta switching:  $ \chi^\textsc{d}(t)=C \delta(t)$}

We consider here the effect of a fast kick of the system, modelled by a delta coupling. This can be seen as the limit of a succession of thinner Gaussian (or top-hat) functions of equal area (recall that this way of interpreting the delta as  a limit is important for the results at hand, as discussed in detail in \cite{Pozas2017}, and in Appendix~\ref{app1}). Due to the pointlike-in-time nature of the interaction, we are explicitly in a regime where the rotating wave approximation does not hold.

\begin{figure*}
	\centering
	\begin{tabular}[c]{cc}
		\begin{tabular}[b]{c}
			\subfloat{\label{fig:right}%
				
				\begin{tabular}[b]{c}
		\includegraphics[scale=.9,valign=l]{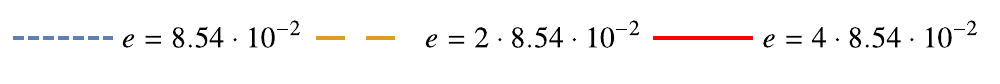}\\[.8mm]
		\hspace*{-4cm}\includegraphics[scale=.78,valign=c]{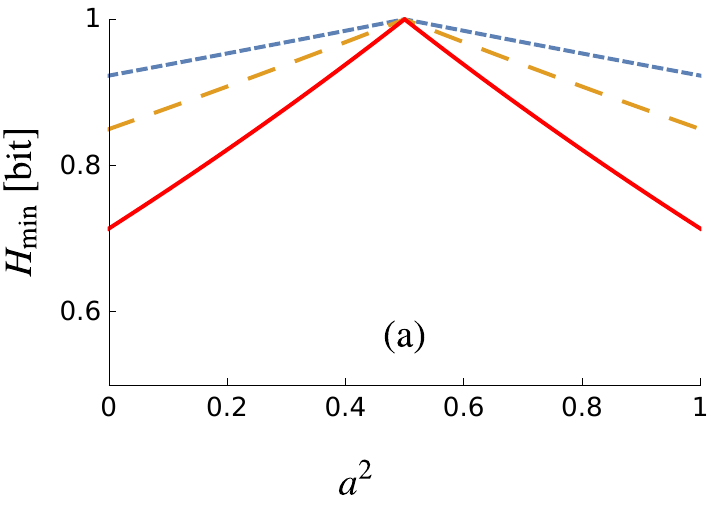}
		
	\end{tabular}}
			
		\end{tabular}
		&
		\subfloat{\label{fig:left}%

			\begin{tabular}[b]{c}
			\includegraphics[scale=.95,valign=c]{leg3.pdf}\\[2mm]
			\hspace*{-3.8cm}	\includegraphics[scale=.79,valign=c]{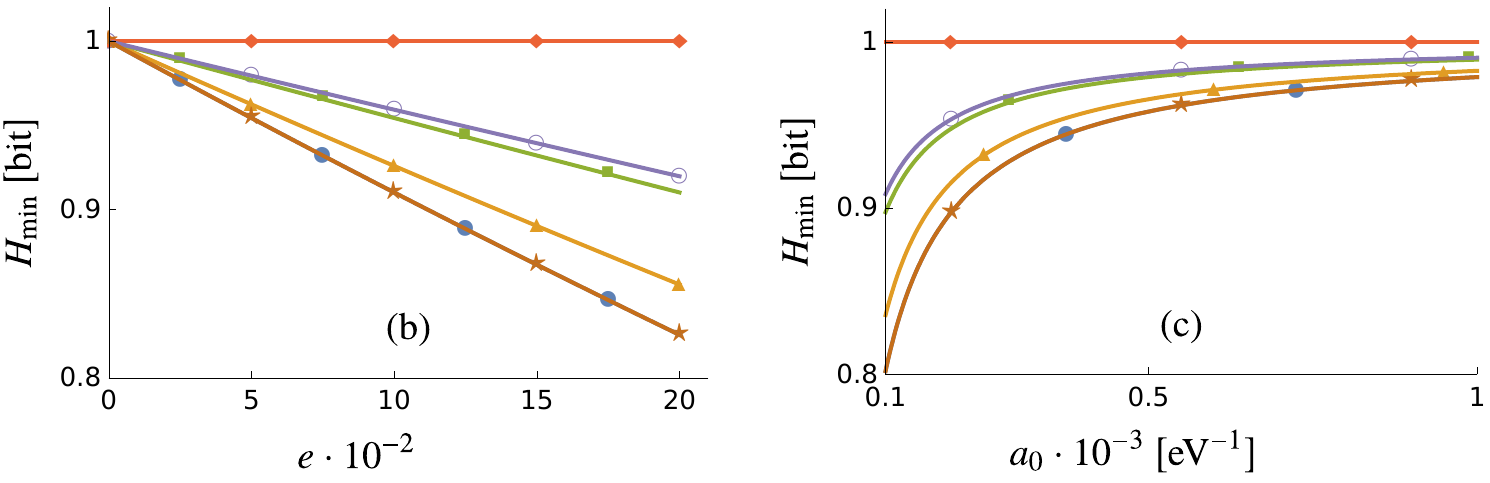}

			\end{tabular}
		}
		
	\end{tabular}
	\caption{(a) Min-entropy $H_{\text{min}}$ plotted against $a^2$ (proportional to the $z$ component of the Bloch vector) with the atomic radius $a_0=2.68\cdot 10^{-4}~\text{eV}^{-1}$ for delta switching and different values of the coupling strength $e$, taking $C=\sigma=2.5 \times 10^{-3}~\text{eV}^{-1}$. At equal superposition $a={1}/\!{\sqrt{2}}$ the extracted randomness has its maximum with $H_{\text{min}}=1~$bit. (b, c) Extracted min-entropy as a function of the parameters (b) electric charge $e$ and (c) atomic radius $a_0$ for different atomic state parameters $a$ in the case of delta switching. In (b) $a_0$ is taken to be $2.68 \cdot10^{-4}~\text{eV}^{-1}$, and in (c) the coupling constant is $8.54\cdot 10^{-2}$.} 
	\label{dirac1}
\end{figure*}

Studying the delta switching, we take $C=\sigma=2.5 \times 10^{-3}~\text{eV}^{-1}$ (reading \eqref{diracfinal} we note that $C$ acts in the same way as the coupling constant). The particular choice for $C$ means that the time-integrated switching function is proportional to $\sigma$, as was in the case for Gaussian and sudden switching. Eq. \!\eqref{diracfinal} shows that once again the equal superposition yields perfect randomness extraction  $H_{\text{min}}=1~\text{bit}$. This can be seen in Fig.~\ref{dirac1}. 

Fig.~\ref{dirac1}a shows that the min-entropy peaks at equal superposition of the atom's initial state and quickly decreases at either side, resulting in a much larger loss of randomness than for any of the previously studied switching functions. 
Moreover the peak becomes narrower the stronger the coupling between atom and field is, spoiling quickly any randomness extraction if it is not in an equal superposition state. 

	\begin{figure*}
		\includegraphics[scale=.7,valign=c]{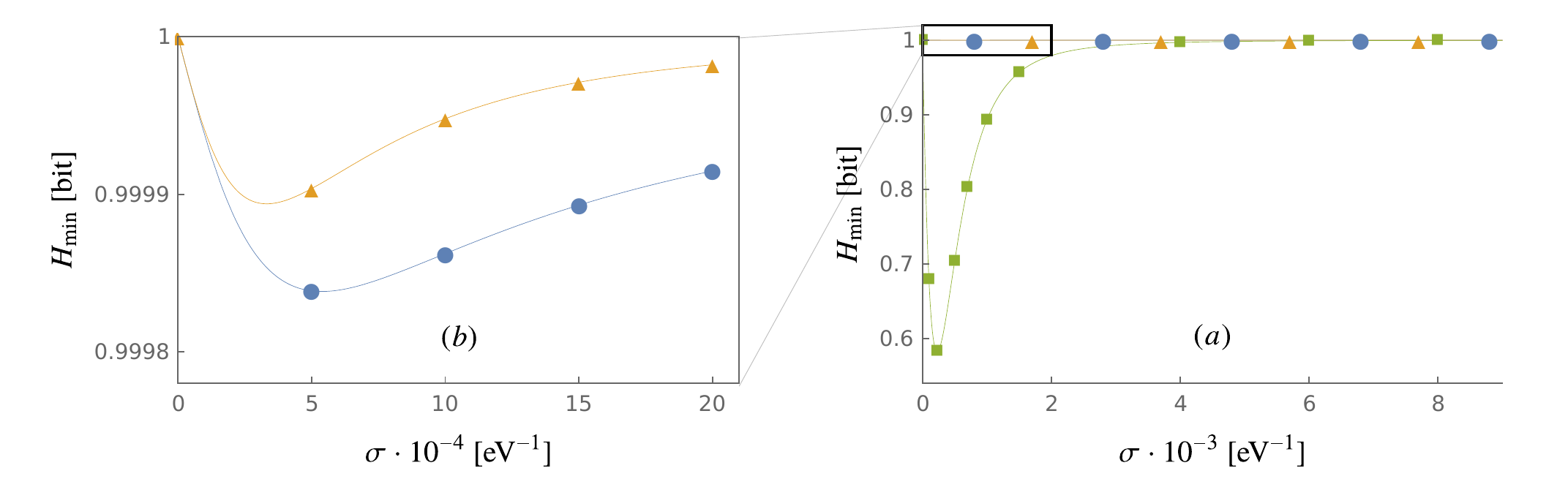}
		\includegraphics[scale=.95,valign=c]{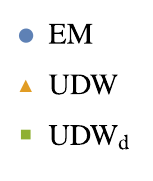}
		\caption{$(a)$ Min-entropy $H_{\text{min}}$ plotted against duration of interaction $\sigma$ with parameters $a_0=2.68 \cdot 10^{-4}~\text{eV}^{-1}$, $e = 10^{-3}$, $\Omega = 3.73~\text{eV}$ in the ground state $a=1$ for Gaussian switching of the electric dipole model (EM)  for $1s \rightarrow 2p_z$, scalar coupling (UDW) and coupling to the time derivative of the scalar field ($\text{UDW}_{\text{d}}$). Both scalar models assume a $1s \rightarrow 2s$ transition. $(b)$ Corresponds to a zoomed-in region of $(a)$ marked by the black box.  The considered interactions are ultra-short, i.e. $\sigma \Omega < 0.03$, hence beyond any RWA.}
		\label{scalar}
	\end{figure*}


From Fig.~\ref{dirac1}b and Fig.~\ref{dirac1}c we see that, as expected, a stronger coupling between atom and electromagnetic field causes larger correlations and reduces the min-entropy. In addition, the dependence on the atomic radius displays an increase to an asymptotic value of the min-entropy. In contrast to the two previous switching functions, the delta switching shows much larger variations in the min-entropy.

	\subsection{Comparison with scalar field models}
	Let us now compare our results to earlier studies where the atom was modeled as an Unruh-DeWitt (UDW) detector coupled to a scalar field $\phi(\bm x, t)$.
	The UDW model has been shown to capture the fundamental features to leading order of light-matter interactions as long as there is no exchange of orbital angular momentum \cite{Pozas2016, Alhambra2014, Montero2013}.
	We will consider two different kinds of UDW detectors, namely the original UDW model \cite{PhysRevD.14.870, deWitt} and the derivative coupling (that we will denote as $\text{UDW}_{\text{d}}$) \cite{derivcoupling}. The respective Hamiltonians are
	\begin{align}
	H_{\text{UDW}}&=e \chi(t) \int \dd \bm x^3 F(\bm x) \hat{\mu}(t) \hat{\phi}(\bm x,t),\\
	H_{\text{UDW}_{\text{d}}}&=e \chi(t) \int \dd \bm x^3 F(\bm x) \hat{\mu}(t) \partial_t \hat{\phi}(\bm x,t) \label{deriv},
	\end{align} 
	where $\hat{\mu}(t)$ is the monopole moment capturing the internal degrees of freedom of the detector, and $F(\bm x)$ is the \textit{ad hoc} included spatial smearing function of the detector. In particular, \eqref{deriv} has been used in previous literature to analyze the loss of randomness due to coupling to relativistic fields \cite{Thinh2016}, so it makes sense to compare the results of the simplified scalar model with the realistic hydrogen-like model employed here. 
	
	The difference between the EM coupling and these two models has been analyzed in the past in the context of entanglement harvesting \cite{Pozas2016}. The $\text{UDW}_{\text{d}}$ model can be thought of as a scalar analogue of the dipole coupling by noting that in the Coulomb gauge $\bm E= - \partial_t \bm A$ and one may perhaps expect that it should resemble the dipole interaction to some extent (as discussed in \cite{Thinh2016}). Both scalar models do not allow transitions where there is exchange of angular momentum. In particular, the  $1s \rightarrow 2p_z$ transition is not permitted. Same as in \cite{Pozas2016} we will consider the closest scalar analogue to that transition to compare to the EM case; that is  $1s \rightarrow 2s$ for the scalar models, but keeping the $1s \rightarrow 2p_z$ transition for the EM model.
	
	The change in the density matrix  of the atomic state after an interaction of time $\sigma$ takes for the scalar couplings the form 
	\begin{align}
	\Delta \hat \rho_{\text{UDW}}=&- \frac{32768}{\pi} \left(a_0^2 e \sigma\right)^2 \!\!\int_0^{\infty}\! \!\dd |\bm k| \frac{|\bm k|^5 e^{-\frac{1}{2}\sigma^2 (|\bm k|+\Omega)^2}}{(4 a_0^2 |\bm k|^2+9)^6}\hat\sigma_z, \\
	\Delta \hat \rho_{\text{UDW}_{\text{d}}}=&- \frac{32768}{\pi} \left(a_0^2 e \sigma\right)^2 \!\!\int_0^{\infty} \!\!\dd |\bm k| \frac{|\bm k|^7 e^{-\frac{1}{2}\sigma^2 (|\bm k|+\Omega)^2}}{(4 a_0^2 |\bm k|^2+9)^6}\hat\sigma_z, \label{derivative}
	\end{align} 
	with $\hat\sigma_z$ being the Pauli Z matrix.  
	The scalar models were derived by assuming a Gaussian switching function and the initial ground state of the detector ($a=1$). In addition the smearing function was chosen as the scalar version of the smearing vector: $F(\bm x)=\psi_e(\bm x) \psi_g(\bm x)$. Consequently, we have to analyze the electric dipole model in the respective configuration slice. It should be noted that the coupling constants $e$ of the different couplings do not all have the same dimensionality. In particular, for the dipole and direct scalar interaction we find $[e]=0$, whilst for the derivative coupling $[e]=-1$ (in mass dimensions). We choose the parameters $a_0=2.68 \cdot 10^{-4}~\text{eV}^{-1}$, $e = 10^{-3}$, $\Omega = 3.73~\text{eV}$, taking into account that for stronger couplings the perturbative expansion of the $\text{UDW}_{\text{d}}$ model breaks down by virtue of the additional $|\bm k|^2$ dependence in \eqref{derivative}.
	
	In Fig.~\ref{scalar} one finds that the derivative model vastly underestimates the extracted randomness for early times and is off by up to over $40~$\%. On the other hand the UDW model slightly overestimates it for short interaction times by the order of $10^{-2}~$\%. For long interaction times both scalar models approach the realistic dipole model. Since $\sigma \Omega \ll 1$ in the plots, we again go beyond the validity of the RWA.

	\section{Conclusions}\label{conclusion}

We quantified a lower bound for the randomness that can be extracted from a hydrogen-like atom coupled to the electromagnetic field. This work is an advancement of the previous work \cite{Thinh2016} by having considered a fully-featured hydrogen-like atom coupling to an electromagnetic field in 3+1 dimensions (instead of a monopole detector coupling to a  scalar field in 1+1D). In doing so we have tackled previous criticisms to former literature in the choice of the smearing function (here derived from the atomic orbital wavefunctions from first principles), the anisotropic nature of atomic transitions and exchange of angular momentum between atom and quantum field, as well as the choice of the value of the physical parameters of the problem (here coming from first principles). Lastly, we compared adiabatic and sudden time-dependencies of the interaction strength. We emphasize that, same as in the simpler models employed in \cite{Thinh2016}, we also did  not make any use of the usual simplifications of the interaction in the context of quantum optics. Namely, we did not assume the rotating wave approximation or the single mode approximation.

We analyzed how much information an adversary with access to the EM field but not the atom can obtain about a supposedly random measurement outcome. We found (consistently with studies that considered simplified scalar field interaction models \cite{Thinh2016}) that generally the ground state of the atom and the vacuum state of the field is not the optimal state to generate randomness out of a succession of preparation and measurement in unbiased bases for the atomic state basis. 

We have analyzed a variety of switching regimes and found that for the switching function as well as the duration of the interaction between atom and electromagnetic field there are two possibilities for choosing the optimal state in terms of randomness generation: For short time between preparation and measurement in the unbiased basis, the equal superposition between ground and excited atomic states yields the optimal randomness. For coupling and decoupling times much shorter than the inverse of the frequency of the atomic transition (that could be thought of as preparation-to-measurement times), the equal superposition between ground and excited yields the best results even for long times between preparation and measurement. In contrast, for adiabatic switching and long times, the ground state of the atom yields the optimal randomness generation. 

Furthermore, we also showed that in the cases where the equal superposition is optimal, the ground state is one of the two worst choices (together with the excited atomic state) in order to generate randomness, something that contradicts the intuition coming from the rotating wave approximation that basically would suggest that `if everything is in the ground state the field and the atom will remain uncrorrelated'.

	Finally, we compared the realistic model of the electromagnetic field coupled to the atom via a dipole moment to simplified scalar models used in previous studies. We found that both the Unruh-Dewitt coupling \cite{deWitt} and the derivative coupling \cite{derivcoupling} provide a good approximation for the full electromagnetic model for long enough interaction times. For short interaction times, the Unruh-Dewitt model is a better approximation than the derivative coupling, which significantly deviates from the full electromagnetic calculation. This information is useful when considering scalar approximations to the light-matter interaction
	
	

We would like to emphasize that the scope of this paper is not as much to describe a particular experimental setup, but rather  study how the  flow of quantum information in special relativistic quantum regimes deviates from non-relativistic scenarios (which make use rotating wave and single mode approximations) by virtue of the entanglement between the atom and the electromagnetic field. In this context we aimed to analyze how this flow of information impacts the ability of obtaining certified randomness, even under ideal assumptions regarding preparation and measurement procedures, as a matter of first principles.

	\section{Acknowledgements}
	E. M-M acknowledges the funding of the NSERC Discovery program and the Ontario Early Researcher Award.
		\begin{widetext}
	\appendix

		\section{Deriving the time-evolved density matrix}\label{app1}
		In this section we will derive in detail the change of the reduced density matrix of the atom after interaction with the electromagnetic field. Starting from Eq. \!\eqref{atomic rho} by recalling Eq. \!\eqref{eq:evo} we find
		\begin{align}
		\tr_{\textsc{f}}\left( \hat U^{(2)} \hat \rho_i\right)=&\tr_{\textsc{f}}\left(-\int_{\mathbb{R}} \dd t \int_{-\infty}^t \dd t' \chi(t) \chi(t') \int_{\mathbb{R}^3} \dd \bm x \int_{\mathbb{R}^3}  \dd \bm x'  \sum_{i,j=1}^3 \hat d_i(\bm x, t)\hat E_i(\bm x, t)\hat d_j(\bm x', t') \hat E_j(\bm x', t') \hat \rho_i \right) \nonumber \\
		=&-\int_{\mathbb{R}} \dd t \int_{-\infty}^t \dd t' \chi(t) \chi(t') \int_{\mathbb{R}^3} \dd \bm x \int_{\mathbb{R}^3}  \dd \bm x'  \sum_{i,j=1}^3 \hat d_i(\bm x, t)  \hat d_j(\bm x', t') \ket \Psi_{\textsc{\!a}}\! \!\bra \Psi      \leftidx{_\textsc{e}\!\!}{\bra 0 \hat E_i(\bm x, t) \hat E_j(\bm x', t') \ket 0}{_\textsc{\!e}}.
		\end{align}
		Since the atomic state outer product $\ket \Psi_{\textsc{\!a}}\! \! \bra \Psi$ is on the right-hand side of the product of the dipole operators, only terms with one raising and one lowering operator survive. Similarly,
		\begin{align}
		\tr_{\textsc{f}}\left( \hat U^{(1)} \hat \rho_i \hat U^{(1) \dagger}\right)=& \int_{\mathbb{R}} \dd t \int_{\mathbb{R}} \dd t' \chi(t) \chi(t') \int_{\mathbb{R}^3} \dd \bm x \int_{\mathbb{R}^3} \dd \bm x' \sum_{i, j=1}^3 \hat d_i(\bm x, t)  \ket \Psi_{\textsc{\!a}}\!\! \bra \Psi \hat d_j(\bm x', t') \tr_{\textsc{f}} \left(\hat  E_i(\bm x, t) \ket 0_{\!\textsc{e}}\!\! \bra 0 \hat E_j(\bm x', t') \right),
		\end{align}
		where we used that $\hat d_i$ and $\hat E_i$ are Hermitian. This yields then
		\begin{align}
		&\tr_{\textsc{f}}\left(\hat  U^{(2)} \hat \rho_i\right) \nonumber\\
		&\begin{aligned}
		= \!-e^2 \!\!\int_{\mathbb{R}} \!\dd t\! \int_{-\infty}^t \!\! \!\dd t' \chi(t) \chi(t') \int_{\mathbb{R}^3} \!\!\dd \bm x \!\int_{\mathbb{R}^3} \! \!\dd \bm x' \!\sum_{i,j=1}^3  \!\!
		& \left\{\left[\left(1-a^2\right) \ket e_\textsc{\!a}\!\! \bra e + a \sqrt{1-a^2} \ket e_\textsc{\!a}\!\! \bra g \right] \right. F_i^T(\bm x)  W_{i j}(\bm x, \bm x'; t, t') F_j^*(\bm x') e^{\ii \Omega(t-t')}   \\
		&~~ \left.   + \left[a^2 \ket g_\textsc{\!a}\!\! \bra g + a \sqrt{1-a^2} \ket g_\textsc{\!a}\!\!\bra e \right]  F_i^{*^T}(\bm x  )  W_{i j}(\bm x, \bm x'; t, t') F_j(\bm x'  ) e^{-\ii \Omega(t-t')} \right\} ,
		\end{aligned} \\
		&\tr_{\textsc{f}}\left( \hat U^{(1)} \hat \rho_i \hat U^{(1) \dagger}\right)\nonumber\\
		&\begin{aligned}
		=& e^2 \int_{\mathbb{R}} \dd t \int_{\mathbb{R}} \dd t' \chi(t) \chi(t') \int_{\mathbb{R}^3} \dd \bm x \int_{\mathbb{R}^3} \dd \bm x' \sum_{i,j=1}^3 \left\{a^2 \ket e_\textsc{\!a}\!\! \bra e  F_i^{*^T}(\bm x'  )  W_{i j}(\bm x', \bm x; t', t)  F_j(\bm x  ) e^{\ii \Omega(t-t')} \right.  \\
		&\quad+ \left(1-a^2\right) \ket g_\textsc{\!a}\!\! \bra g  F_i^{T}(\bm x'  ) W_{i j}(\bm x', \bm x; t', t) F_j^*(\bm x  ) e^{-\ii \Omega(t-t')}     + a \sqrt{1-a^2} \ket g_\textsc{\!a}\!\!\bra e  F_i^{*^T}(\bm x'  ) W_{i j}(\bm x', \bm x; t', t) F_j^*(\bm x  ) e^{-\ii \Omega(t+t')}  \\
		&\quad+ \left. a \sqrt{1-a^2} \ket e_\textsc{\!a}\!\! \bra g  F_i^{T}(\bm x'  ) W_{i j}(\bm x', \bm x; t', t) F_j(\bm x  ) e^{\ii \Omega(t+t')} \right\} ,
		\end{aligned} 
		\end{align}
		where we have defined the Wightman 2-tensor for the electric field
		\begin{equation}
		W_{i j}(\bm x_2, \bm x_1; t_2, t_1)= \leftidx{_\textsc{e}\!\!}{ \bra 0 \hat{ E}_i(\bm x_2, t_2) \hat{ E}_j(\bm x_1, t_1) \ket 0}{_\textsc{\!e}}= \int_{\mathbb{R}^3}  \frac{\dd^3 \bm{k}}{(2 \pi)^{3}} \frac{|\bm k|}{2}e^{-\ii |\bm k| (t_2-t_1)} e^{\ii \bm k \cdot (\bm x_2 - \bm x_1)} \left( \delta_{i, j} - \frac{ k_i  k_j}{|\bm k|^2}\right).
		\end{equation}
		To arrive at that expression the completeness relation of the polarization vectors $ \bm \epsilon(\bm k, s_i)$ was used:
		\begin{align}
		\sum_{i=1}^2  \bm \epsilon(\bm k, s_i) \otimes \bm \epsilon(\bm k, s_i) =  \openone - \frac{\bm k \otimes \bm k}{|\bm k|^2}.
		\end{align}
		In the following we will drop the subscripts of the outer products belonging to the Hilbert space of atom A. We will separate the terms in the Wightman tensor according to the identity $\openone$ and the dyadic product $\bm k \otimes \bm k$ (such that their sum corresponds to the complete expression), denoted by the corresponding subscripts.
		We wish to integrate over spherical coordinates, naturally suggested by the wave function $\Psi_{n l m}(\bm x)= R_{n l}(|\bm x|) Y_{l m }(\hat{\bm x})$, where $Y_{l m }(\hat{\bm x})$ are the spherical harmonics with $\hat{\bm x}=(\theta_{\bm x}, \phi_{\bm x})$ as the angular components of the unit radial vector, and $R_{n l}(|\bm x|)$ are the radial wave functions of a hydrogenoid atom \cite{demtroeder}. The following two decompositions are helpful:
		\begin{align}
		e^{\ii \bm x \cdot \bm y}&= \sum_{l=0}^\infty \sum_{m=-l}^l 4 \pi \ii^l j_l(|\bm x| |\bm y|) Y_{l m}(\hat{\bm x})Y^*_{l m}(\hat{\bm y})= \sum_{l=0}^\infty \sum_{m=-l}^l 4 \pi \ii^l j_l(|\bm x| |\bm y|) Y^*_{l m}(\hat{\bm x})Y_{l m}(\hat{\bm y}), \\
		\bm x\cdot\bm y&=\frac{4 \pi}{3} |\bm x| |\bm y| \left[ Y_{1 0 }(\hat{\bm x})Y_{1 0}(\hat{\bm y})-Y_{1 1 }(\hat{\bm x})Y_{1 -1}(\hat{\bm y})-Y_{1 -1 }(\hat{\bm x})Y_{1 1}(\hat{\bm y}) \right],
		\end{align}
		with the spherical Bessel functions $j_l(x)$.
		The first contribution to the time evolved density matrix then reads
		\begin{align}
		&\left. \tr_{\textsc{f}}\left(\hat  U^{(2)} \hat \rho_i\right)\right|_{\openone} \nonumber \\
		&= -e^2 \int_{0}^{\infty} \frac{\dd |\bm k|}{(2 \pi)^3} \frac{|\bm k|^3}{2} \sum_{l=0}^\infty \sum_{m=-l}^l 4 \pi \ii^l \sum_{l'=0}^\infty \sum_{m'=-l'}^{l'} 4 \pi \ii^{l'} (-1)^{l'} \frac{4 \pi}{3} \int_{\mathbb{R}} \dd t \int_{-\infty}^t \dd t' \chi(t) \chi(t') e^{-\ii |\bm k| (t-t')}  \nonumber \\ 
		&\quad \times\int_{0}^{\infty} \dd |\bm x| |\bm x|^3 R_{n_e l_e}(|\bm x|)R_{n_g l_g}(|\bm x|) j_l(|\bm k| |\bm x|) \int_{0}^{\infty} \dd |\bm x'| |\bm x'|^3 R_{n_e l_e}(|\bm x'|)R_{n_g l_g}(|\bm x'|) j_{l'}(|\bm k| |\bm x'|)\nonumber \\
		&\quad \times\int \dd \Omega_k  Y_{l m}(\hat{\bm k})Y_{l' m'}(\hat{\bm k})  \int \dd \Omega_{x} \int \dd \Omega_{x'} Y_{l m}^*(\hat{\bm x}) Y_{l' m'}^*(\hat{\bm x'}) \left[ Y_{1 0 }(\hat{\bm x})Y_{1 0}(\hat{\bm x'})-Y_{1 1 }(\hat{\bm x})Y_{1 -1}(\hat{\bm x'})-Y_{1 -1 }(\hat{\bm x})Y_{1 1}(\hat{\bm x'}) \right] \nonumber \\
		&\quad\times \left\{ \left[\left(1-a^2\right) \ket e \! \bra e + a \sqrt{1-a^2} \ket e \! \bra g \right] e^{\ii \Omega(t-t')} Y^*_{l_e m_e}(\hat{\bm x}) Y_{l_g m_g}(\hat{\bm x}) Y_{l_e m_e}(\hat{\bm x'}) Y^*_{l_g m_g}(\hat{\bm x'})\right. \nonumber \\
		&\quad \quad  \left. + \left[ a^2 \ket g \! \bra g + a \sqrt{1-a^2} \ket g \!\bra e \right] e^{-\ii \Omega (t-t')} Y_{l_e m_e}(\hat{\bm x}) Y^*_{l_g m_g}(\hat{\bm x}) Y^*_{l_e m_e}(\hat{\bm x'}) Y_{l_g m_g}(\hat{\bm x'}) \right\} \label{traceU2II},
		\end{align}
		where we have used the identity $Y_{l m}(-\hat{\bm x})=(-1)^l Y_{l m}(\hat{\bm x})$ and that $R_{n l}(|\bm x|)$ is real. Also $\text{d}\Omega=\text{d}(\cos{\Theta}) \text{d}\phi$ is the standard solid angle differential.
		The integral over $\dd \Omega_k$ reads $\int \dd \Omega_k Y_{l m}(\hat{\bm k})Y_{l' m'}(\hat{\bm k})= (-1)^{m'} \delta_{l, l'} \delta_{m, -m'}$ by using $Y^*_{l m}(\hat{\bm x})=(-1)^m Y_{l -m}(\hat{\bm x})$. This simplifies the integrals over the other two solid angles drastically such that we can use the following identity of spherical harmonics integrated over the unit sphere $S^2$
		\begin{align}
		&\int \dd\Omega~ Y_{l_1, m_1}^*(\hat{\bm x}) Y_{l_3, m_3}^*(\hat{\bm x}) Y_{l_2, m_2}(\hat{\bm x}) Y_{l_4, m_4}(\hat{\bm x})\nonumber \\
		&= \sum\limits_{\lambda=0}^{\infty} \sum\limits_{\mu=-\lambda}^{\lambda} \frac{2 \lambda+1}{4 \pi} \sqrt{(2l_1+1)(2l_2+1)(2l_3+1)(2l_4+1)} \begin{pmatrix}
		l_1 & l_3 &\lambda \\
		0&0&0\\
		\end{pmatrix}
		\begin{pmatrix}
		l_2 & l_4 &\lambda \\
		0&0&0\\
		\end{pmatrix}
		\begin{pmatrix}
		l_1 & l_3 &\lambda \\
		-m_1&-m_3&-\mu\\
		\end{pmatrix}
		\begin{pmatrix}
		l_2 & l_4 &\lambda \\
		m_2&m_4&\mu\\
		\end{pmatrix},
		\end{align}
		with $\begin{pmatrix}
		l_1 & l_2 &l_3 \\
		m_1&m_2&m_3\\
		\end{pmatrix}$
		as the Wigner 3$j$-symbols (see, for instance, section 34.2 of \cite{NIST:DLMF}). With this formula, the sums over $l', m, m'$ and the integrals over the all solid angles can be executed. Let us concentrate first on the second term of the sum in the curly brackets of \eqref{traceU2II}, coming from $ F_i^{*^T}  W_{i j} F_j$, which yields
		\begin{align}
		& \sum_{l'=0}^{\infty}\sum_{m=-l}^{l}\sum_{m'=-l'}^{l'} \ii^{l+l'} (-1)^{l'} j_{l'}(|\bm k| |\bm x'|)\int \dd \Omega_k  Y_{l m}(\hat{\bm k})Y_{l' m'}(\hat{\bm k})  \int \dd \Omega_{x} Y_{l_e m_e}(\hat{\bm x}) Y^*_{l_g m_g}(\hat{\bm x}) Y_{l m}^*(\hat{\bm x}) 
		\nonumber \\ 
		&\times\int \dd \Omega_{x'}   Y^*_{l_e m_e}(\hat{\bm x'}) Y_{l_g m_g}(\hat{\bm x'})  Y_{l' m'}^*(\hat{\bm x'}) \left[ Y_{1 0 }(\hat{\bm x})Y_{1 0}(\hat{\bm x'})-Y_{1 1 }(\hat{\bm x})Y_{1 -1}(\hat{\bm x'})-Y_{1 -1 }(\hat{\bm x})Y_{1 1}(\hat{\bm x'}) \right] \nonumber \\
		&= \frac{3 (-1)^{m_g-m_e} \ii^{2 l} (-1)^l}{(4 \pi)^2}(2l+1)(2l_e+1)(2l_g+1)\sum_{\lambda,\lambda'=0}^{\infty} (2\lambda+1)(2\lambda'+1) j_l(|\bm k| |\bm x'|) \begin{pmatrix}
		l & l_g &\lambda \\
		0&0&0\\
		\end{pmatrix}
		\begin{pmatrix}
		l_e & 1 &\lambda \\
		0&0&0\\
		\end{pmatrix}
		\nonumber \\
		&  \quad\times  \begin{pmatrix}
		l & l_e &\lambda' \\
		0&0&0\\
		\end{pmatrix}
		\begin{pmatrix}
		l_g & 1 &\lambda' \\
		0&0&0\\
		\end{pmatrix} \left[  \begin{pmatrix}
		l & l_g &\lambda \\
		m_g-m_e&-m_g&m_e\\
		\end{pmatrix}
		\begin{pmatrix}
		l_e & 1 &\lambda \\
		m_e&0&-m_e\\
		\end{pmatrix}
		\begin{pmatrix}
		l & l_e &\lambda' \\
		m_e-m_g&-m_e&m_g\\
		\end{pmatrix}
		\begin{pmatrix}
		l_g & 1 &\lambda' \\
		m_g&0&-m_g\\
		\end{pmatrix} \right. \nonumber \\
		& \quad\quad +\begin{pmatrix}
		l & l_g &\lambda \\
		1+m_g-m_e&-m_g&m_e-1\\
		\end{pmatrix}
		\begin{pmatrix}
		l_e & 1 &\lambda \\
		m_e&-1&1-m_e\\
		\end{pmatrix}
		\begin{pmatrix}
		l & l_e &\lambda' \\
		m_e-m_g-1&-m_e&1+m_g\\
		\end{pmatrix}
		\begin{pmatrix}
		l_g & 1 &\lambda' \\
		m_g&1&-1-m_g\\
		\end{pmatrix}\nonumber \\
		& \quad \quad +\left. \begin{pmatrix}
		l & l_g &\lambda \\
		m_g-m_e-1&-m_g&m_e+1\\
		\end{pmatrix}
		\begin{pmatrix}
		l_e & 1 &\lambda \\
		m_e&1&-1-m_e\\
		\end{pmatrix}
		\begin{pmatrix}
		l & l_e &\lambda' \\
		1+m_e-m_g&-m_e&m_g-1\\
		\end{pmatrix}
		\begin{pmatrix}
		l_g & 1 &\lambda' \\
		m_g&-1&1-m_g\\
		\end{pmatrix} \right], \label{3sum}
		\end{align}
		where the sums over $\mu$ and $\mu'$ can be executed by using that 3$j$-symbols are zero unless the sum over the entries of the lower row is zero.
		The first term from \eqref{traceU2II} can be obtained from \eqref{3sum} by noting that effectively $l$ and $l'$, thus also $m$ and $m'$, are interchanged and hence it requires to take $(-1)^{m_g-m_e} \rightarrow (-1)^{-m_g+m_e}$. Since this is equivalent, the first term can also be described by \eqref{3sum}. Therefore, in all generality \eqref{traceU2II} reads
		\begin{align}
		&\left. \tr_{\textsc{f}}\left(\hat  U^{(2)} \hat \rho_i\right)\right|_{\openone}\nonumber \\
		&= -e^2 \int_{0}^{\infty} \frac{\dd |\bm k|}{(2 \pi)^3} \frac{|\bm k|^3}{2} \sum_{l=0}^\infty (4 \pi)^2  \frac{4 \pi}{3} \int_{\mathbb{R}} \dd t \int_{-\infty}^t \dd t' \chi(t) \chi(t') e^{-\ii |\bm k| (t-t')} \int_{0}^{\infty} \dd |\bm x| |\bm x|^3 R_{n_e l_e}(|\bm x|)R_{n_g l_g}(|\bm x|) j_l(|\bm k| |\bm x|) \nonumber \\
		&\quad\times\!\int_{0}^{\infty}\! \!\!\dd |\bm x'| |\bm x'|^3 R_{n_e l_e}(|\bm x'|)R_{n_g l_g}(|\bm x'|) j_{l}(|\bm k| |\bm x'|)\frac{3 (-1)^{m_g-m_e}}{(4 \pi)^2}(2l+1)(2l_e+1)(2l_g+1)\!\!\sum_{\lambda,\lambda'=0}^{\infty}\!\! \!(2\lambda+1)(2\lambda'+1) 
		\! \begin{pmatrix}
		l & l_g &\lambda \\
		0&0&0\\
		\end{pmatrix} \nonumber \\
		&\quad	\times \begin{pmatrix}
		l_e & 1 &\lambda \\
		0&0&0\\
		\end{pmatrix} 
		\!\begin{pmatrix}
		l & l_e &\lambda' \\
		0&0&0\\
		\end{pmatrix}
		\!\begin{pmatrix}
		l_g & 1 &\lambda' \\
		0&0&0\\
		\end{pmatrix} \!\nonumber \left[ \! \begin{pmatrix}
		l & l_g &\lambda \\
		m_g-m_e&-m_g&m_e\\
		\end{pmatrix}
		\!\begin{pmatrix}
		l_e & 1 &\lambda \\
		m_e&0&-m_e\\
		\end{pmatrix}
		\!\begin{pmatrix}
		l & l_e &\lambda' \\
		m_e-m_g&-m_e&m_g\\
		\end{pmatrix}
		\!\begin{pmatrix}
		l_g & 1 &\lambda' \\
		m_g&0&-m_g\\
		\end{pmatrix}\! \right. \nonumber \\
		&\quad\quad+\begin{pmatrix}
		l & l_g &\lambda \\
		1+m_g-m_e&-m_g&m_e-1\\
		\end{pmatrix}
		\begin{pmatrix}
		l_e & 1 &\lambda \\
		m_e&-1&1-m_e\\
		\end{pmatrix}
		\begin{pmatrix}
		l & l_e &\lambda' \\
		m_e-m_g-1&-m_e&1+m_g\\
		\end{pmatrix}
		\begin{pmatrix}
		l_g & 1 &\lambda' \\
		m_g&1&-1-m_g\\
		\end{pmatrix}\nonumber \\
		&\quad\quad+ \left. \begin{pmatrix}
		l & l_g &\lambda \\
		m_g-m_e-1&-m_g&m_e+1\\
		\end{pmatrix}
		\begin{pmatrix}
		l_e & 1 &\lambda \\
		m_e&1&-1-m_e\\
		\end{pmatrix}
		\begin{pmatrix}
		l & l_e &\lambda' \\
		1+m_e-m_g&-m_e&m_g-1\\
		\end{pmatrix}
		\begin{pmatrix}
		l_g & 1 &\lambda' \\
		m_g&-1&1-m_g\\
		\end{pmatrix} \right] \nonumber\\
		&\quad\times\left\{ \left[\left(1-a^2\right) \ket e \! \bra e + a \sqrt{1-a^2} \ket e \! \bra g \right] e^{\ii \Omega(t-t')}+\left[ a^2 \ket g \! \bra g + a \sqrt{1-a^2} \ket g \!\bra e \right] e^{-\ii \Omega (t-t')} \right\}.
		\end{align}
		Before specifying atomic transition or the switching function of the coupling to the electric field, we will derive the general expressions of the remaining terms, having derived terms containing  $F_i^{*^T} W_{i j} F_j$ and $F_i^T W_{i j} F_j^*$ of the $\openone$ part. Secondly we look at the remaining $\openone$ contribution:
		\begin{align}
		&\left. \tr_{\textsc{f}}\left( \hat U^{(1)} \hat \rho_i \hat U^{(1) \dagger}\right)\right|_{\openone}\nonumber \\
		&= e^2 \int_{0}^{\infty} \frac{\dd |\bm k|}{(2 \pi)^3} \frac{|\bm k|^3}{2} \sum_{l=0}^\infty \sum_{m=-l}^l 4 \pi \ii^l \sum_{l'=0}^\infty \sum_{m'=-l'}^{l'} 4 \pi \ii^{l'} (-1)^{l'} \frac{4 \pi}{3} \int_{\mathbb{R}} \dd t \int_{\mathbb{R}} \dd t' \chi(t) \chi(t') e^{-\ii |\bm k| (t'-t)} \int \dd \Omega_k  Y_{l m}(\hat{\bm k})Y_{l' m'}(\hat{\bm k})  \nonumber \\ 
		&\quad \times\!\int_{0}^{\infty}\!\!\! \dd |\bm x| |\bm x|^3 R_{n_e l_e}(|\bm x|)R_{n_g l_g}(|\bm x|) j_l(|\bm k| |\bm x|) \!\int_{0}^{\infty}\!\!\! \dd |\bm x'| |\bm x'|^3 R_{n_e l_e}(|\bm x'|)R_{n_g l_g}(|\bm x'|) j_{l'}(|\bm k| |\bm x'|)\! \!\int \!\! \dd \Omega_{x}\! \!\int \!\!\dd \Omega_{x'} Y_{l m}^*(\hat{\bm x}) Y_{l' m'}^*(\hat{\bm x'})\nonumber \\
		&\quad \times\! \left[ Y_{1 0 }(\hat{\bm x})Y_{1 0}(\hat{\bm x'})\!-\!Y_{1 1 }(\hat{\bm x})Y_{1 -1}(\hat{\bm x'})\!-\!Y_{1 -1 }(\hat{\bm x})Y_{1 1}(\hat{\bm x'}) \right]\!  \left\{\left(1-a^2\right) \!\ket g \! \bra g \!e^{-\ii \Omega(t-t')} Y^*_{l_e m_e}(\hat{\bm x'}) Y_{l_g m_g}(\hat{\bm x'}) Y_{l_e m_e}(\hat{\bm x}) Y^*_{l_g m_g}(\hat{\bm x})\right.\nonumber \\
		&\quad\quad +  a^2 \ket e \! \bra e e^{\ii \Omega (t-t')} Y_{l_e m_e}(\hat{\bm x'}) Y^*_{l_g m_g}(\hat{\bm x'}) Y^*_{l_e m_e}(\hat{\bm x}) Y_{l_g m_g}(\hat{\bm x}) +  a \sqrt{1-a^2} \ket g \! \bra e e^{-\ii \Omega (t+t')} Y_{l_e m_e}(\hat{\bm x'}) Y^*_{l_g m_g}(\hat{\bm x'}) Y_{l_e m_e}(\hat{\bm x})\nonumber \\
		&\quad\quad\left. \times Y^*_{l_g m_g}(\hat{\bm x})+  a \sqrt{1-a^2} \ket e \! \bra g e^{\ii \Omega (t+t')} Y^*_{l_e m_e}(\hat{\bm x'}) Y_{l_g m_g}(\hat{\bm x'}) Y^*_{l_e m_e}(\hat{\bm x}) Y_{l_g m_g}(\hat{\bm x}) \right\} \label{traceU1II}.
		\end{align}
		From \eqref{3sum} we already know how to compute the first two terms in the curly brackets. The other two follow immediately by noting that they can be obtained from the known terms by including or removing the conjugate of one of the smearing functions. Either way, effectively $l_e \leftrightarrow l_g$ and $m_e \leftrightarrow m_g$ change in the corresponding term. As we recall from \eqref{3sum}, we had the requirement that $m=-m'$ and hence all contributions disappear except when $m_e=m_g$ since the Wigner 3-j symbols vanish in case the sum of the lower components does not equal zero.
		Thus we find
		\begin{align}
		&\left. \tr_{\textsc{f}}\left(  \hat U^{(1)} \hat \rho_i \hat U^{(1) \dagger}\right)\right|_{\openone} \nonumber \\
		&= e^2 \int_{0}^{\infty} \frac{\dd |\bm k|}{(2 \pi)^3} \frac{|\bm k|^3}{2} \sum_{l=0}^\infty (4 \pi)^2  \frac{4 \pi}{3} \int_{\mathbb{R}} \dd t \int_{\mathbb{R}} \dd t' \chi(t) \chi(t') e^{-\ii |\bm k| (t'-t)} \int_{0}^{\infty} \dd |\bm x| |\bm x|^3 R_{n_e l_e}(|\bm x|)R_{n_g l_g}(|\bm x|) j_l(|\bm k| |\bm x|)\nonumber \\
		&\quad \times\int_{0}^{\infty} \dd |\bm x'| |\bm x'|^3 R_{n_e l_e}(|\bm x'|)R_{n_g l_g}(|\bm x'|) j_{l}(|\bm k| |\bm x'|) \frac{3 (-1)^{m_g-m_e}}{(4 \pi)^2}(2l+1)(2l_e+1)(2l_g+1)\sum_{\lambda,\lambda'=0}^{\infty} (2\lambda+1)(2\lambda'+1) \nonumber \\ 
		&\quad\times \left\{ \left(\left(1-a^2\right) \ket g \! \bra g e^{-\ii \Omega(t-t')} + a^2 \ket e \! \bra e e^{\ii \Omega (t-t')} \right)
		\begin{pmatrix}
		l & l_g &\lambda \\
		0&0&0\\
		\end{pmatrix}
		\begin{pmatrix}
		l_e & 1 &\lambda \\
		0&0&0\\
		\end{pmatrix}
		\begin{pmatrix}
		l & l_e &\lambda' \\
		0&0&0\\
		\end{pmatrix}
		\begin{pmatrix}
		l_g & 1 &\lambda' \\
		0&0&0\\
		\end{pmatrix}  \right. \nonumber \\
		&\quad\quad\times \left[  \begin{pmatrix}
		l & l_g &\lambda \\
		m_g-m_e&-m_g&m_e\\
		\end{pmatrix}
		\begin{pmatrix}
		l_e & 1 &\lambda \\
		m_e&0&-m_e\\
		\end{pmatrix}
		\begin{pmatrix}
		l & l_e &\lambda' \\
		m_e-m_g&-m_e&m_g\\
		\end{pmatrix}
		\begin{pmatrix}
		l_g & 1 &\lambda' \\
		m_g&0&-m_g\\
		\end{pmatrix} \right. \nonumber \\
		& \quad \quad\quad+\begin{pmatrix}
		l & l_g &\lambda \\
		1+m_g-m_e&-m_g&m_e-1\\
		\end{pmatrix}
		\begin{pmatrix}
		l_e & 1 &\lambda \\
		m_e&-1&1-m_e\\
		\end{pmatrix}
		\begin{pmatrix}
		l & l_e &\lambda' \\
		m_e-m_g-1&-m_e&1+m_g\\
		\end{pmatrix}
		\begin{pmatrix}
		l_g & 1 &\lambda' \\
		m_g&1&-1-m_g\\
		\end{pmatrix}\nonumber \\
		&\quad\quad \quad + \left. \begin{pmatrix}
		l & l_g &\lambda \\
		m_g-m_e-1&-m_g&m_e+1\\
		\end{pmatrix}
		\begin{pmatrix}
		l_e & 1 &\lambda \\
		m_e&1&-1-m_e\\
		\end{pmatrix}
		\begin{pmatrix}
		l & l_e &\lambda' \\
		1+m_e-m_g&-m_e&m_g-1\\
		\end{pmatrix}
		\begin{pmatrix}
		l_g & 1 &\lambda' \\
		m_g&-1&1-m_g\\
		\end{pmatrix} \right] \nonumber\\
		& \quad\quad +  \delta_{m_e , m_g}  a \sqrt{1-a^2}   \ket e \! \bra g e^{\ii \Omega (t+t')} \begin{pmatrix}
		l & l_e &\lambda \\
		0&0&0\\
		\end{pmatrix}
		\begin{pmatrix}
		l_g &1 &\lambda \\
		0&0&0\\
		\end{pmatrix}
		\begin{pmatrix}
		l & l_e &\lambda' \\
		0&0&0\\
		\end{pmatrix}
		\begin{pmatrix}
		l_g & 1 &\lambda' \\
		0&0&0\\
		\end{pmatrix} \nonumber \\
		&\quad \quad\times \left[  \begin{pmatrix}
		l & l_e &\lambda \\
		m_e-m_g&-m_e&m_g\\
		\end{pmatrix}
		\begin{pmatrix}
		l_g & 1 &\lambda \\
		m_g&0&-mg\\
		\end{pmatrix}
		\begin{pmatrix}
		l & l_e &\lambda' \\
		m_e-m_g&-m_e&m_g\\
		\end{pmatrix}
		\begin{pmatrix}
		l_g & 1 &\lambda' \\
		m_g&0&-m_g\\
		\end{pmatrix} \right. \nonumber \\
		&\quad\quad \quad +\begin{pmatrix}
		l & l_e &\lambda \\
		1+m_e-m_g&-m_e&m_g-1\\
		\end{pmatrix}
		\begin{pmatrix}
		l_g & 1 &\lambda \\
		m_g&-1&1-mg\\
		\end{pmatrix}
		\begin{pmatrix}
		l & l_e &\lambda' \\
		m_e-m_g-1&-m_e&1+m_g\\
		\end{pmatrix}
		\begin{pmatrix}
		l_g & 1 &\lambda' \\
		m_g&1&-1-m_g\\
		\end{pmatrix}\nonumber \\
		&\quad \quad \quad +\left. \begin{pmatrix}
		l & l_e &\lambda \\
		m_e-m_g-1&-m_e&m_g+1\\
		\end{pmatrix}
		\begin{pmatrix}
		l_g & 1 &\lambda \\
		m_g&1&-1-mg\\
		\end{pmatrix}
		\begin{pmatrix}
		l & l_e &\lambda' \\
		1+m_e-m_g&-m_e&m_g-1\\
		\end{pmatrix}
		\begin{pmatrix}
		l_g & 1 &\lambda' \\
		m_g&-1&1-m_g\\
		\end{pmatrix} \right] \nonumber \\
		&\quad \quad +  \delta_{m_e , m_g}  a \sqrt{1-a^2}   \ket g \! \bra e e^{-\ii \Omega (t+t')} \begin{pmatrix}
		l & l_g &\lambda \\
		0&0&0\\
		\end{pmatrix}
		\begin{pmatrix}
		l_e & 1 &\lambda \\
		0&0&0\\
		\end{pmatrix}
		\begin{pmatrix}
		l & l_g &\lambda' \\
		0&0&0\\
		\end{pmatrix}
		\begin{pmatrix}
		l_e & 1 &\lambda' \\
		0&0&0\\
		\end{pmatrix} \nonumber \\
		&\quad \quad \times\left[  \begin{pmatrix}
		l & l_g &\lambda \\
		m_g-m_e&-m_g&m_e\\
		\end{pmatrix}
		\begin{pmatrix}
		l_e & 1 &\lambda \\
		m_e&0&-m_e\\
		\end{pmatrix}
		\begin{pmatrix}
		l & l_g &\lambda' \\
		m_g-m_e&-m_g&m_e\\
		\end{pmatrix}
		\begin{pmatrix}
		l_e & 1 &\lambda' \\
		m_e&0&-m_e\\
		\end{pmatrix} \right. \nonumber \\
		&\quad\quad \quad +\begin{pmatrix}
		l & l_g &\lambda \\
		1+m_g-m_e&-m_g&m_e-1\\
		\end{pmatrix}
		\begin{pmatrix}
		l_e & 1 &\lambda \\
		m_e&-1&1-m_e\\
		\end{pmatrix}
		\begin{pmatrix}
		l & l_g &\lambda' \\
		m_g-m_e-1&-m_g&1+m_e\\
		\end{pmatrix}
		\begin{pmatrix}
		l_e & 1 &\lambda' \\
		m_e&1&-1-m_e\\
		\end{pmatrix}\nonumber \\
		&\left.\quad \quad \quad +\left. \begin{pmatrix}
		l & l_g &\lambda \\
		m_g-m_e-1&-m_g&m_e+1\\
		\end{pmatrix}
		\begin{pmatrix}
		l_e & 1 &\lambda \\
		m_e&1&-1-m_e\\
		\end{pmatrix}
		\begin{pmatrix}
		l & l_g &\lambda' \\
		1+m_g-m_e&-m_g&m_e-1\\
		\end{pmatrix}
		\begin{pmatrix}
		l_e & 1 &\lambda' \\
		m_e&-1&1-m_e\\
		\end{pmatrix} \right] \right\},
		\end{align}
		where we explicitly used the Kronecker delta to indicate the dependence on $m_e$ and $m_g$. One can derive the second from the first term in the curly brackets by exchanging $l_e \leftrightarrow l_g$ in terms associated to $\lambda$, and derive the third from the first term by exchanging $l_e \leftrightarrow l_g$ in terms associated to $\lambda'$.
		
		Now we can determine the formulae for the dyadic part $\bm k \otimes \bm k$. We start from
		\begin{align}
		& \sum_{m=-l}^{l}\sum_{m'=-l'}^{l'} \int \dd \Omega_k  Y_{l m}(\hat{\bm k})Y_{l' m'}(\hat{\bm k})  \int \dd \Omega_{x} Y_{l_e m_e}(\hat{\bm x}) Y^*_{l_g m_g}(\hat{\bm x}) Y_{l m}^*(\hat{\bm x}) \int \dd \Omega_{x'}   Y^*_{l_e m_e}(\hat{\bm x'}) Y_{l_g m_g}(\hat{\bm x'})  Y_{l' m'}^*(\hat{\bm x'}) \nonumber \\
		&\times \left[ Y_{1 0 }(\hat{\bm x})Y_{1 0}(\hat{\bm k})-Y_{1 1 }(\hat{\bm x})Y_{1 -1}(\hat{\bm k})-Y_{1 -1 }(\hat{\bm x})Y_{1 1}(\hat{\bm k}) \right] \left[ Y_{1 0 }(\hat{\bm k})Y_{1 0}(\hat{\bm x'})-Y_{1 1 }(\hat{\bm k})Y_{1 -1}(\hat{\bm x'})-Y_{1 -1 }(\hat{\bm k})Y_{1 1}(\hat{\bm x'}) \right] \nonumber \\
		&= 9 (2l+1)(2l'+1)(2l_g+1)(2l_e+1)\! \sum_{\lambda', \lambda''=0}^{\infty} \!\frac{(2\lambda'+1)(2\lambda''+1)}{(4\pi)^3}\! \begin{pmatrix}
		l & l_g &\lambda' \\
		0&0&0\\
		\end{pmatrix}
		\!\begin{pmatrix}
		l_e & 1 &\lambda' \\
		0&0&0\\
		\end{pmatrix}
		\! \begin{pmatrix}
		l' & l_e &\lambda'' \\
		0&0&0\\
		\end{pmatrix}
		\!\begin{pmatrix}
		l_g & 1 &\lambda'' \\
		0&0&0\\
		\end{pmatrix}
		\!(A+B),
		\end{align}
		with 
		\begin{align}
		A=&\sum_{\lambda=0}^\infty (2\lambda+1)
		\tj{1}{1}{\lambda}{0}{0}{0}^2
		\tj{l}{l'}{\lambda}{0}{0}{0}\tj{l}{l_g}{\lambda'}{m_g-m_e}{-m_g}{m_e}
		\tj{l_e}{1}{\lambda'}{m_e}{0}{-m_e}\notag\\
		&\quad\times\tj{l}{l'}{\lambda}{m_e-m_g}{m_g-m_e}{0}
		\tj{l'}{l_e}{\lambda''}{m_e-m_g}{-m_e}{m_g}
		\tj{l_g}{1}{\lambda''}{m_g}{0}{-m_g}\notag\\
		&+\sum_{\lambda=0}^\infty(2\lambda+1)
		\tj{1}{1}{\lambda}{0}{0}{0}
		\tj{1}{1}{\lambda}{0}{1}{-1}
		\tj{l}{l'}{\lambda}{0}{0}{0}
		\notag\\
		&\quad\times\left[\tj{l}{l'}{\lambda}{m_e-m_g-1}{m_g-m_e}{1}
		\tj{l'}{l_e}{\lambda''}{m_e-m_g}{-m_e}{m_g}
		\tj{l_g}{1}{\lambda''}{m_g}{0}{-m_g}\right.\notag\\
		&\qquad\times\tj{l}{l_g}{\lambda'}{m_g-m_e+1}{-m_g}{m_e-1}\tj{l_e}{1}{\lambda'}{m_e}{-1}{1-m_e}\notag\\
		&\quad+\tj{l}{l'}{\lambda}{m_e-m_g}{m_g-m_e-1}{1}
		\tj{l'}{l_e}{\lambda''}{m_e-m_g+1}{-m_e}{m_g-1}
		\tj{l_g}{1}{\lambda''}{m_g}{-1}{1-m_g}\notag\\
		&\qquad\times\left.\tj{l}{l_g}{\lambda'}{m_g-m_e}{-m_g}{m_e}\tj{l_e}{1}{\lambda'}{m_e}{0}{-m_e}\right]\notag\\
		&+\sum_{\lambda=0}^\infty(2\lambda+1)
		\tj{1}{1}{\lambda}{0}{0}{0}
		\tj{1}{1}{\lambda}{0}{-1}{1}
		\tj{l}{l'}{\lambda}{0}{0}{0}
		\notag\\
		&\quad\times\left[\tj{l}{l'}{\lambda}{m_e-m_g+1}{m_g-m_e}{-1}
		\tj{l'}{l_e}{\lambda''}{m_e-m_g}{-m_e}{m_g}
		\tj{l_g}{1}{\lambda''}{m_g}{0}{-m_g}\right.\notag\\
		&\qquad\times\tj{l}{l_g}{\lambda'}{m_g-m_e-1}{-m_g}{m_e+1}
		\tj{l_e}{1}{\lambda'}{m_e}{1}{-1-m_e}\notag\\
		&\quad+\tj{l}{l'}{\lambda}{m_e-m_g}{m_g-m_e+1}{-1}
		\tj{l'}{l_e}{\lambda''}{m_e-m_g-1}{-m_e}{m_g+1}
		\tj{l_g}{1}{\lambda''}{m_g}{1}{-1-m_g}\notag\\
		&\left.\qquad\times\tj{l}{l_g}{\lambda'}{m_g-m_e}{-m_g}{m_e}
		\tj{l_e}{1}{\lambda'}{m_e}{0}{-m_e}\right]\notag\\
		&+\sum_{\lambda=0}^\infty(2\lambda+1)
		\tj{1}{1}{\lambda}{0}{0}{0}
		\tj{1}{1}{\lambda}{1}{-1}{0}
		\tj{l}{l'}{\lambda}{0}{0}{0}
		\notag\\
		&\quad\times
		\left[\tj{l}{l'}{\lambda}{m_e-m_g-1}{m_g-m_e+1}{0}
		\tj{l'}{l_e}{\lambda''}{m_e-m_g-1}{-m_e}{m_g+1}
		\tj{l_g}{1}{\lambda''}{m_g}{1}{-1-m_g}\right.\notag\\
		&\qquad\times\tj{l}{l_g}{\lambda'}{m_g-m_e+1}{-m_g}{m_e-1}
		\tj{l_e}{1}{\lambda'}{m_e}{-1}{1-m_e}\notag\\
		&\quad+\tj{l}{l'}{\lambda}{m_e-m_g+1}{m_g-m_e-1}{0}
		\tj{l'}{l_e}{\lambda''}{m_e-m_g+1}{-m_e}{m_g-1}
		\tj{l_g}{1}{\lambda''}{m_g}{-1}{1-m_g}\notag\\
		&\left.\qquad\times\tj{l}{l_g}{\lambda'}{m_g-m_e-1}{-m_g}{m_e+1}
		\tj{l_e}{1}{\lambda'}{m_e}{1}{-1-m_e}\right].
		\end{align}
		
		\begin{align}
		B=&\sqrt{\frac{2}{3}}
		\tj{l}{l'}{2}{0}{0}{0}
		\tj{l}{l'}{2}{m_e-m_g-1}{m_g-m_e-1}{2}
		\tj{l'}{l_e}{\lambda''}{1+m_e-m_g}{-m_e}{m_g-1}
		\notag\\
		&\quad\times\tj{l_g}{1}{\lambda''}{m_g}{-1}{1-m_g}
		\tj{l_e}{1}{\lambda'}{m_e}{-1}{1-m_e}
		\tj{l}{l_g}{\lambda'}{1-m_e+m_g}{-m_g}{m_e-1}\notag\\
		&+\sqrt{\frac{2}{3}}
		\tj{l}{l'}{2}{0}{0}{0}
		\tj{l}{l'}{2}{m_e-m_g+1}{m_g-m_e+1}{-2}
		\tj{l'}{l_e}{\lambda''}{m_e-m_g-1}{-m_e}{m_g+1}\notag\\
		&\quad\times\tj{l_g}{1}{\lambda''}{m_g}{1}{-1-m_g}
		\tj{l}{l_g}{\lambda'}{m_g-m_e-1}{-m_g}{m_e+1}
		\tj{l_e}{1}{\lambda'}{m_e}{1}{-1-m_e}.
		\end{align}
		This yields hence
		\begin{align}
		&\left. \tr_{\textsc{f}}\left(\hat  U^{(2)} \hat \rho_i\right)\right|_{\bm k \otimes \bm k}\nonumber \\
		&= e^2 \int_{0}^{\infty} \frac{\dd |\bm k|}{(2 \pi)^3} \frac{|\bm k|^3}{2} \sum_{l=0}^\infty \sum_{m=-l}^l 4 \pi \ii^l \sum_{l'=0}^\infty \sum_{m'=-l'}^{l'} 4 \pi \ii^{l'} (-1)^{l'} \left(\frac{4 \pi}{3}\right)^2 \int_{\mathbb{R}} \dd t \int_{-\infty}^t \dd t' \chi(t) \chi(t') e^{-\ii |\bm k| (t-t')}  \nonumber \\ 
		&\quad\times \int_{0}^{\infty} \dd |\bm x| |\bm x|^3 R_{n_e l_e}(|\bm x|)R_{n_g l_g}(|\bm x|) j_l(|\bm k| |\bm x|) \int_{0}^{\infty} \dd |\bm x'| |\bm x'|^3 R_{n_e l_e}(|\bm x'|)R_{n_g l_g}(|\bm x'|) j_{l'}(|\bm k| |\bm x'|)\nonumber \\
		&\quad \times\int \dd \Omega_k  Y_{l m}(\hat{\bm k})Y_{l' m'}(\hat{\bm k})  \int \dd \Omega_{x} \int \dd \Omega_{x'} Y_{l m}^*(\hat{\bm x}) Y_{l' m'}^*(\hat{\bm x'}) \left[ Y_{1 0 }(\hat{\bm x})Y_{1 0}(\hat{\bm k})-Y_{1 1 }(\hat{\bm x})Y_{1 -1}(\hat{\bm k})-Y_{1 -1 }(\hat{\bm x})Y_{1 1}(\hat{\bm k}) \right]\nonumber\\
		& \quad\times  \left[ Y_{1 0 }(\hat{\bm k})Y_{1 0}(\hat{\bm x'})-Y_{1 1 }(\hat{\bm k})Y_{1 -1}(\hat{\bm x'})-Y_{1 -1 }(\hat{\bm k})Y_{1 1}(\hat{\bm x'}) \right] \nonumber \\
		&\quad\times \left\{ \left[\left(1-a^2\right) \ket e \! \bra e + a \sqrt{1-a^2} \ket e \! \bra g \right] e^{\ii \Omega(t-t')} Y^*_{l_e m_e}(\hat{\bm x}) Y_{l_g m_g}(\hat{\bm x}) Y_{l_e m_e}(\hat{\bm x'}) Y^*_{l_g m_g}(\hat{\bm x'})\right. \nonumber \\
		&\quad\quad \left. + \left[ a^2 \ket g \! \bra g + a \sqrt{1-a^2} \ket g \! \bra e \right] e^{-\ii \Omega (t-t')} Y_{l_e m_e}(\hat{\bm x}) Y^*_{l_g m_g}(\hat{\bm x}) Y^*_{l_e m_e}(\hat{\bm x'}) Y_{l_g m_g}(\hat{\bm x'}) \right\} \nonumber \\
		&=  e^2 \int_{0}^{\infty} \frac{\dd |\bm k|}{(2 \pi)^3} \frac{|\bm k|^3}{2} 4 \pi \int_{\mathbb{R}} \dd t \int_{-\infty}^t \dd t' \chi(t) \chi(t') e^{-\ii |\bm k| (t-t')}  \sum_{l,l'=0}^\infty \ii^{l+l'}(2l+1)(2l'+1)(2l_g+1)(2l_e+1)\nonumber \\
		&\quad \times \int_{0}^{\infty} \dd |\bm x| |\bm x|^3 R_{n_e l_e}(|\bm x|)R_{n_g l_g}(|\bm x|) \int_{0}^{\infty} \dd |\bm x'| |\bm x'|^3 R_{n_e l_e}(|\bm x'|)R_{n_g l_g}(|\bm x'|)\nonumber \\
		&\quad\times \sum_{\lambda', \lambda''=0}^{\infty}(2\lambda'+1)(2\lambda''+1) \begin{pmatrix}
		l & l_g &\lambda' \\
		0&0&0\\
		\end{pmatrix}
		\begin{pmatrix}
		l_e & 1 &\lambda' \\
		0&0&0\\
		\end{pmatrix}
		\begin{pmatrix}
		l' & l_e &\lambda'' \\
		0&0&0\\
		\end{pmatrix}
		\begin{pmatrix}
		l_g & 1 &\lambda'' \\
		0&0&0\\
		\end{pmatrix}
		(A+B) \nonumber\\
		&\quad\times \left\{(-1)^{l}  j_l(|\bm k| |\bm x'|) j_{l'}(|\bm k| |\bm x|) \left[\left(1-a^2\right) \ket e \! \bra e + a \sqrt{1-a^2} \ket e \! \bra g \right] e^{\ii \Omega(t-t')} \right. \nonumber \\
		&\quad\quad \left. + (-1)^{l'}  j_l(|\bm k| |\bm x|) j_{l'}(|\bm k| |\bm x'|) \left[ a^2 \ket g \! \bra g + a \sqrt{1-a^2} \ket g \! \bra e \right] e^{-\ii \Omega (t-t')} \right\}, \label{traceU2kk}
		\end{align}
		where we redefined $l \leftrightarrow l'$ for the first term in the curly brackets to derive the last formula.
		
		The last contribution we have to calculate is analogously
		\begin{align}
		&\left. \tr_{\textsc{f}}\left( \hat U^{(1)} \hat \rho_i \hat U^{(1) \dagger}\right)\right|_{\bm k \otimes \bm k}\nonumber \\
		&=- e^2 \int_{0}^{\infty} \frac{\dd |\bm k|}{(2 \pi)^3} \frac{|\bm k|^3}{2} \sum_{l=0}^\infty \sum_{m=-l}^l 4 \pi \ii^l \sum_{l'=0}^\infty \sum_{m'=-l'}^{l'} 4 \pi \ii^{l'} (-1)^{l'} \left(\frac{4 \pi}{3}\right)^2 \int_{\mathbb{R}} \dd t \int_{\mathbb{R}} \dd t' \chi(t) \chi(t') e^{-\ii |\bm k| (t'-t)}  \nonumber \\ 
		&\quad \times\int_{0}^{\infty} \dd |\bm x| |\bm x|^3 R_{n_e l_e}(|\bm x|)R_{n_g l_g}(|\bm x|) j_l(|\bm k| |\bm x|) \int_{0}^{\infty} \dd |\bm x'| |\bm x'|^3 R_{n_e l_e}(|\bm x'|)R_{n_g l_g}(|\bm x'|) j_{l'}(|\bm k| |\bm x'|)\nonumber \\
		&\quad\times \int \dd \Omega_k  Y_{l m}(\hat{\bm k})Y_{l' m'}(\hat{\bm k})  \int \dd \Omega_{x} \int \dd \Omega_{x'} Y_{l m}^*(\hat{\bm x}) Y_{l' m'}^*(\hat{\bm x'})  \left[ Y_{1 0 }(\hat{\bm x})Y_{1 0}(\hat{\bm k})-Y_{1 1 }(\hat{\bm x})Y_{1 -1}(\hat{\bm k})-Y_{1 -1 }(\hat{\bm x})Y_{1 1}(\hat{\bm k}) \right]\nonumber\\
		&\quad\times \!\left[ Y_{1 0 }(\hat{\bm k})Y_{1 0}(\hat{\bm x'})\!-\!Y_{1 1 }(\hat{\bm k})Y_{1 -1}(\hat{\bm x'})\!-\!Y_{1 -1 }(\hat{\bm k})Y_{1 1}(\hat{\bm x'}) \right] \!\left\{\!\left(1-a^2\right) \ket g \! \bra g e^{-\ii \Omega(t-t')} Y^*_{l_e m_e}(\hat{\bm x'}) Y_{l_g m_g}(\hat{\bm x'}) Y_{l_e m_e}(\hat{\bm x}) Y^*_{l_g m_g}(\hat{\bm x})\right. \nonumber \\
		&\quad\quad+  a^2 \ket e \! \bra e e^{\ii \Omega (t-t')} Y_{l_e m_e}(\hat{\bm x'}) Y^*_{l_g m_g}(\hat{\bm x'}) Y^*_{l_e m_e}(\hat{\bm x}) Y_{l_g m_g}(\hat{\bm x}) \nonumber \\
		&\quad\quad+  a \sqrt{1-a^2} \ket g \! \bra e e^{-\ii \Omega (t+t')} Y_{l_e m_e}(\hat{\bm x'}) Y^*_{l_g m_g}(\hat{\bm x'}) Y_{l_e m_e}(\hat{\bm x}) Y^*_{l_g m_g}(\hat{\bm x}) \nonumber \\
		&\quad\quad\left. +  a \sqrt{1-a^2} \ket e \! \bra g e^{\ii \Omega (t+t')} Y^*_{l_e m_e}(\hat{\bm x'}) Y_{l_g m_g}(\hat{\bm x'}) Y^*_{l_e m_e}(\hat{\bm x}) Y_{l_g m_g}(\hat{\bm x}) \right\}\nonumber \\
		&=- e^2 \!\int_{0}^{\infty}\!\! \frac{\dd |\bm k|}{(2 \pi)^3} \frac{|\bm k|^3}{2}  4 \pi \!\int_{\mathbb{R}} \!\!\dd t \!\!\int_{\mathbb{R}} \!\dd t' \chi(t) \chi(t') e^{-\ii |\bm k| (t'-t)} \! \!\int_{0}^{\infty}\! \!\!\dd |\bm x| |\bm x|^3 R_{n_e l_e}(|\bm x|)R_{n_g l_g}(|\bm x|)\! \!\int_{0}^{\infty}\!\! \!\dd |\bm x'| |\bm x'|^3 R_{n_e l_e}(|\bm x'|)R_{n_g l_g}(|\bm x'|)\nonumber \\
		&\quad\times\! \!\sum_{l,l'=0}^{\infty}\! \ii^{l+l'} (2l+1)(2l'+1)(2l_g+1)(2l_e+1)\! \!\sum_{\lambda', \lambda''=0}^{\infty}\!\! (2\lambda'+1)(2\lambda''+1) \left\{
		\! \begin{pmatrix}
		l & l_g &\lambda' \\
		0&0&0\\
		\end{pmatrix}
		\begin{pmatrix}
		l_e & 1 &\lambda' \\
		0&0&0\\
		\end{pmatrix}
		\! \begin{pmatrix}
		l' & l_e &\lambda'' \\
		0&0&0\\
		\end{pmatrix}
		\!\begin{pmatrix}
		l_g & 1 &\lambda'' \\
		0&0&0\\
		\end{pmatrix}
		\right. \nonumber\\
		&\quad\quad \times (A+B)  \left[(-1)^{l}  j_{l'}(|\bm k| |\bm x'|) j_{l}(|\bm k| |\bm x|) \left(1-a^2\right) \ket g \! \bra g e^{-\ii \Omega(t-t')} + (-1)^{l'}  j_{l}(|\bm k| |\bm x'|) j_{l'}(|\bm k| |\bm x|) a^2 \ket e \! \bra e e^{\ii \Omega (t-t')} \right] \nonumber \\
		&\quad\quad + \!\delta_{m_e}^{m_g}\!\begin{pmatrix}
		l & l_g &\lambda' \\
		0&0&0\\
		\end{pmatrix}
		\!\begin{pmatrix}
		l_e & 1 &\lambda' \\
		0&0&0\\
		\end{pmatrix}
		\!\begin{pmatrix}
		l' & l_g &\lambda'' \\
		0&0&0\\
		\end{pmatrix}
		\!\begin{pmatrix}
		l_e & 1 &\lambda'' \\
		0&0&0\\
		\end{pmatrix}
		\!\left(\tilde A+\tilde B\right)   (-1)^{l'}  j_{l}(|\bm k| |\bm x'|) j_{l'}(|\bm k| |\bm x|)  a \sqrt{1-a^2} \ket g \! \bra e e^{-\ii \Omega (t+t')}  \nonumber \\
		&\left.\quad\quad+\!\delta_{m_e}^{m_g}\! \begin{pmatrix}
		l & l_e &\lambda' \\
		0&0&0\\
		\end{pmatrix}
		\!\begin{pmatrix}
		l_g & 1 &\lambda' \\
		0&0&0\\
		\end{pmatrix}
		\! \begin{pmatrix}
		l' & l_e &\lambda'' \\
		0&0&0\\
		\end{pmatrix}
		\!\begin{pmatrix}
		l_g & 1 &\lambda'' \\
		0&0&0\\
		\end{pmatrix}
		\!\left(\tilde{\tilde  A}+\tilde{\tilde B}\right)   (-1)^{l'}  j_{l}(|\bm k| |\bm x'|) j_{l'}(|\bm k| |\bm x|)  a \sqrt{1-a^2} \ket e \! \bra g e^{\ii \Omega (t+t')}  \right\}, \label{traceU1kk}
		\end{align}
		with $\tilde{x}$ indicating $l_e \leftrightarrow l_g$ ($m_e \leftrightarrow m_g$) in $3j$-symbols involving $\lambda''$ (therefore also for the $m$-component of $l'$) and $\tilde{\tilde{x}}$ that $l_e \leftrightarrow l_g$ ($m_e \leftrightarrow m_g$) in $3j$-symbols involving $\lambda'$ (also for the $m$-component of $l$). For instance, 
		\begin{align*}
		x=&\tj{l}{l'}{2}{m_e-m_g-1}{m_g-m_e-1}{2}\tj{l_g}{1}{\lambda''}{m_g}{-1}{1-m_g}
		\tj{l_e}{1}{\lambda'}{m_e}{-1}{1-m_e},\\
		\tilde{x}=&\tj{l}{l'}{2}{m_e-m_g-1}{m_e-m_g-1}{2}\tj{l_e}{1}{\lambda''}{m_e}{-1}{1-m_e}
		\tj{l_e}{1}{\lambda'}{m_e}{-1}{1-m_e},\\
		\tilde{\tilde{x}}=&\tj{l}{l'}{2}{m_g-m_e-1}{m_g-m_e-1}{2}\tj{l_g}{1}{\lambda''}{m_g}{-1}{1-m_g}
		\tj{l_g}{1}{\lambda'}{m_g}{-1}{1-m_g}.
		\end{align*}

		This implies that the coefficients contain terms of the form
		\begin{align}
		\begin{pmatrix}
		l & l' &\lambda \\
		m_1&m_2&m_3\\
		\end{pmatrix}, ~ m_1+m_2+m_3 \neq 0.
		\end{align}
		Therefore, unless $m_e=m_g$, the third and last term of the curly brackets vanish in \eqref{traceU1kk}. The Kronecker delta has been explicitly added to stress this fact.

		Now all expressions are in generality and cannot be simplified more without specifying the atomic transition and the switching function.
		
		\subsection{Transition from ground to first excited state}\label{partic}
		In the following we will derive the time evolved density matrix by studying the $1s \rightarrow 2p_z$ transition ($l_g=0$, $m_g=0$, $l_e=1$, $m_e=1$). Then \eqref{traceU2II} can be simplified by using the properties of the Wigner $3j$-symbols. In particular the first $3j$-symbol forces $l=\lambda$, moreover we see that $\lambda'=1$ and $\lambda=0,1,2$. Thus it yields
		\begin{align}
		&\left. \tr_{\textsc{f}}\left(\hat  U^{(2)} \hat \rho_i\right)\right|_{\openone}\nonumber \\*
		&= -e^2 \int_{0}^{\infty} \frac{\dd |\bm k|}{(2 \pi)^3} \frac{|\bm k|^3}{2} 4 \pi \int_{\mathbb{R}} \dd t \int_{-\infty}^t \dd t' \chi(t) \chi(t') e^{-\ii |\bm k| (t-t')}  \int_{0}^{\infty} \dd |\bm x| |\bm x|^3 R_{2, 1}(|\bm x|)R_{1,0}(|\bm x|) \nonumber \\ 
		&\quad\times \int_{0}^{\infty} \dd |\bm x'| |\bm x'|^3 R_{2, 1}(|\bm x'|)R_{1,0}(|\bm x'|)  \frac{1}{3} \left[ j_0(|\bm k| |\bm x|)j_0(|\bm k| |\bm x'|)+ 2  j_2(|\bm k| |\bm x|)j_2(|\bm k| |\bm x'|) \right] \nonumber  \\
		&\quad\times\left\{ \left[\left(1-a^2\right) \ket e \! \bra e + a \sqrt{1-a^2} \ket e \! \bra g \right] e^{\ii \Omega(t-t')}+\left[ a^2 \ket g \! \bra g + a \sqrt{1-a^2} \ket g \! \bra e \right] e^{-\ii \Omega (t-t')} \right\}.
		\end{align}
		We solve the integral over $|\bm x|$ and $|\bm x'|$ by using the following identity
		\begin{align}
		\int_0^\infty\text{d}r\, r^3R_{2,1}(r)R_{1,0}(r)j_l(|\bm k|r)=8 \sqrt{2 \pi } 3^{-l-\frac{11}{2}} a_0^{l+1} |\bm k|^l \Gamma (l+5) \, _2\tilde{F}_1\left(\frac{l+5}{2},\frac{l+6}{2};l+\frac{3}{2};-\frac{4}{9}  a_0^2 |\bm k|^2\right),
		\label{spaceintegral}
		\end{align}
		with $_2\tilde{F}_1\left(a,b;c;z\right)\coloneqq{}_2F_1(a,b;c;z)/\Gamma(z)$ as the regularized hypergeometric function. Therefore we find
		\begin{align}
		\left. \tr_{\textsc{f}}\left(\hat  U^{(2)} \hat \rho_i\right)\right|_{\openone}=& -e^2 a_0^2\frac{663552}{\pi^2} \int_{0}^{\infty} \dd |\bm k| |\bm k|^3 \int_{\mathbb{R}} \dd t \int_{-\infty}^t \dd t' \chi(t) \chi(t') e^{-\ii |\bm k| (t-t')} \frac{16 a_0^4 |\bm k|^4-8 a_0^2 |\bm k|^2+9}{\left(4 a_0^2 |\bm k|^2+9\right)^8} \nonumber \\ 
		&\times\left\{ \left[\left(1-a^2\right) \ket e \! \bra e + a \sqrt{1-a^2} \ket e \! \bra g \right] e^{\ii \Omega(t-t')}+\left[ a^2 \ket g \! \bra g + a \sqrt{1-a^2} \ket g \! \bra e \right] e^{-\ii \Omega (t-t')} \right\}.
		\end{align}
		Before specifying the switching function $\chi(t)$ to integrate over the time integrals, we will simplify the other contributions to the time evolved density matrix. In particular we find analogously for \eqref{traceU1II} 
		\begin{align}
		&\left. \tr_{\textsc{f}}\left(  \hat U^{(1)} \hat \rho_i \hat U^{(1) \dagger}\right)\right|_{\openone} \nonumber \\
		&= e^2 \int_{0}^{\infty} \frac{\dd |\bm k|}{(2 \pi)^3} \frac{|\bm k|^3}{2}4 \pi \int_{\mathbb{R}} \dd t \int_{\mathbb{R}} \dd t' \chi(t) \chi(t') e^{-\ii |\bm k| (t'-t)}  \int_{0}^{\infty} \dd |\bm x| |\bm x|^3 R_{2, 1}(|\bm x|)R_{1, 0}(|\bm x|) j_l(|\bm k| |\bm x|) \nonumber \\ 
		&\quad\times \int_{0}^{\infty} \dd |\bm x'| |\bm x'|^3 R_{2, 1}(|\bm x'|)R_{1, 0}(|\bm x'|) j_{l}(|\bm k| |\bm x'|) \frac{1}{3} \left[ j_0(|\bm k| |\bm x|)j_0(|\bm k| |\bm x'|)+ 2  j_2(|\bm k| |\bm x|)j_2(|\bm k| |\bm x'|) \right] \nonumber  \\
		& \quad\times\left\{\left(1-a^2\right) \ket g \! \bra g e^{-\ii \Omega(t-t')} + a^2 \ket e \! \bra e e^{\ii \Omega (t-t')} + a \sqrt{1-a^2}   \ket e \! \bra g e^{\ii \Omega (t+t')}  +   a \sqrt{1-a^2}   \ket g \! \bra e e^{-\ii \Omega (t+t')} \right\},
		\end{align} 
		where we used for last term in the curly brackets that $\lambda=\lambda'=l$ and $\lambda=0, 1, 2$. For the penultimate term in the curly brackets one finds $\lambda=\lambda'=1$ and $l=0, 1, 2$. By virtue of \eqref{spaceintegral} we arrive at
		\begin{align}
		&\left. \tr_{\textsc{f}}\left(  \hat U^{(1)} \hat \rho_i \hat U^{(1) \dagger}\right)\right|_{\openone}\nonumber \\
		&= e^2 a_0^2 \frac{663552}{\pi^2} \int_{0}^{\infty}\dd |\bm k||\bm k|^3 \int_{\mathbb{R}} \dd t \int_{\mathbb{R}} \dd t' \chi(t) \chi(t') e^{-\ii |\bm k| (t'-t)} \frac{16 a_0^4 |\bm k|^4-8 a_0^2 |\bm k|^2+9}{\left(4 a_0^2 |\bm k|^2+9\right)^8}  \nonumber \\ 
		&\quad \times\left\{\left(1-a^2\right) \ket g \! \bra g e^{-\ii \Omega(t-t')} + a^2 \ket e \! \bra e e^{\ii \Omega (t-t')} + a \sqrt{1-a^2}   \ket e \! \bra g e^{\ii \Omega (t+t')}  +   a \sqrt{1-a^2}   \ket g \! \bra e e^{-\ii \Omega (t+t')} \right\}.
		\end{align}
		The same will be shown now for the dyadic contributions. Starting from \eqref{traceU2kk} it yields
		\begin{align}
		&\left. \tr_{\textsc{f}}\left(\hat  U^{(2)} \hat \rho_i\right)\right|_{\bm k \otimes \bm k}\nonumber \\
		&=  e^2 \int_{0}^{\infty} \frac{\dd |\bm k|}{(2 \pi)^3} \frac{|\bm k|^3}{2} 4 \pi \int_{\mathbb{R}} \dd t \int_{-\infty}^t \dd t' \chi(t) \chi(t') e^{-\ii |\bm k| (t-t')}\int_{0}^{\infty}\dd |\bm x| |\bm x|^3 R_{2, 1}(|\bm x|)R_{1, 0}(|\bm x|)\nonumber\\
		&\quad \times \int_{0}^{\infty}\dd |\bm x'| |\bm x'|^3 R_{2, 1}(|\bm x'|)R_{1, 0}(|\bm x'|) \frac{1}{9} \left[ j_0(|\bm k| |\bm x|)- 2 j_2(|\bm k| |\bm x|)\right]\left[ j_0(|\bm k| |\bm x'|)-2 j_2(|\bm k| |\bm x'|)\right]  \nonumber \\
		&\quad \times \left\{ \left[\left(1-a^2\right) \ket e \! \bra e + a \sqrt{1-a^2} \ket e \! \bra g \right] e^{\ii \Omega(t-t')} +  \left[ a^2 \ket g \! \bra g + a \sqrt{1-a^2} \ket g \! \bra e \right] e^{-\ii \Omega (t-t')} \right\}, 
		\end{align}
		by noting that $\lambda=\lambda'=1$ and $l$, $l'$, $\lambda=0, 1, 2$. Solving again the spatial integrals we obtain the form
		\begin{align}
		\left. \tr_{\textsc{f}}\left(\hat  U^{(2)} \hat \rho_i\right)\right|_{\bm k \otimes \bm k}&=  e^2 a_0^2 \frac{24576}{\pi^2}\int_{0}^{\infty} \dd |\bm k| |\bm k|^3 \int_{\mathbb{R}} \dd t \int_{-\infty}^t \dd t' \chi(t) \chi(t') e^{-\ii |\bm k| (t-t')} \frac{\left(9-20 a_0^2 |\bm k|^2\right)^2}{\left(4 a_0^2 |\bm k|^2 +9\right)^8}\nonumber\\
		&\quad\times \left\{ \left[\left(1-a^2\right) \ket e \! \bra e + a \sqrt{1-a^2} \ket e \! \bra g \right] e^{\ii \Omega(t-t')} +  \left[ a^2 \ket g \! \bra g + a \sqrt{1-a^2} \ket g \! \bra e \right] e^{-\ii \Omega (t-t')} \right\}.
		\end{align}
		Finally, \eqref{traceU1kk} gives
		\begin{align}
		&\left. \tr_{\textsc{f}}\left( \hat U^{(1)} \hat \rho_i \hat U^{(1) \dagger}\right)\right|_{\bm k \otimes \bm k} \nonumber \\
		&=- e^2 \!\int_{0}^{\infty} \!\frac{\dd |\bm k|}{(2 \pi)^3} \frac{|\bm k|^3}{2}  4 \pi \!\int_{\mathbb{R}}\! \dd t \!\int_{\mathbb{R}}\! \dd t' \chi(t) \chi(t') e^{-\ii |\bm k| (t'-t)} \int_{0}^{\infty} \dd |\bm x| |\bm x|^3 R_{2, 1}(|\bm x|)R_{1, 0}(|\bm x|)  \int_{0}^{\infty} \dd |\bm x'| |\bm x'|^3 R_{2, 1}(|\bm x'|)R_{1, 0}(|\bm x'|)\nonumber \\
		&\quad \times\frac{1}{9} \left[ j_0(|\bm k| |\bm x|)- 2 j_2(|\bm k| |\bm x|)\right] \left[j_0(|\bm k| |\bm x'|)-2 j_2(|\bm k| |\bm x'|)\right]  \nonumber \\
		&\quad\times\left\{ \left(1-a^2\right) \ket g \! \bra g e^{-\ii \Omega(t-t')} +  a^2 \ket e \! \bra e e^{\ii \Omega (t-t')} +  a \sqrt{1-a^2} \ket g \! \bra e e^{-\ii \Omega (t+t')} +  a \sqrt{1-a^2} \ket e \! \bra g e^{\ii \Omega (t+t')}  \right\} \nonumber \\
		&=- e^2 a_0^2 \frac{24576}{\pi^2}\int_{0}^{\infty} \dd |\bm k| |\bm k|^3  \int_{\mathbb{R}} \dd t \int_{\mathbb{R}} \dd t' \chi(t) \chi(t') e^{-\ii |\bm k| (t'-t)}  \frac{\left(9-20 a_0^2 |\bm k|^2\right)^2}{\left(4 a_0^2 |\bm k|^2 +9\right)^8} \nonumber \\ 
		&\quad\times\left\{ \left(1-a^2\right) \ket g \! \bra g e^{-\ii \Omega(t-t')} +  a^2 \ket e \! \bra e e^{\ii \Omega (t-t')} +  a \sqrt{1-a^2} \ket g \! \bra e e^{-\ii \Omega (t+t')} +  a \sqrt{1-a^2} \ket e \! \bra g e^{\ii \Omega (t+t')}  \right\}.
		\end{align}
		
		Now that we have found the analytic expressions evaluated except for the wave vector and time integrals, we can particularize to the desired switching functions and execute the remaining integrals.
		
		\subsubsection{Gaussian, sudden, and delta switching}

		Here we present the results for three different switching functions: $(i)~ \chi^\textsc{g}(t)=e^{-t^2/\sigma^2}$
		\begin{align}
		&\left. \tr_{\textsc{f}}\left( \hat U^{(1)} \hat \rho_i \hat U^{(1) \dagger}\right)\right|_{\openone} \nonumber \\
		&\begin{aligned}
		= \frac{663552}{\pi} (e a_0 \sigma)^2 \int_0^{\infty} \dd|\bm k| |\bm k|^3 \frac{16 a_0^4 |\bm k|^4-8 a_0^2 |\bm k|^2+9 }{(4 a_0^2 |\bm k|^2+9)^8}
		\begin{pmatrix}
		(1-a^2) e^{-\frac{1}{2} \sigma^2(|\bm k|-\Omega)^2} & a \sqrt{1-a^2} e^{-\frac{1}{2}\sigma^2(|\bm k|^2+\Omega^2)} \\
		a\sqrt{1-a^2} e^{-\frac{1}{2}\sigma^2(|\bm k|^2+\Omega^2)}&  a^2 e^{-\frac{1}{2} \sigma^2(|\bm k|+\Omega)^2} \\
		\end{pmatrix}, 
		\end{aligned}\\
		&\left.\tr_{\textsc{f}}\left( \hat U^{(1)} \hat \rho_i \hat U^{(1) \dagger}\right)\right|_{\bm k \otimes \bm k}\nonumber \\
		&\begin{aligned}
		= -\frac{24576}{\pi} (e a_0 \sigma)^2 \int_0^{\infty} \dd|\bm k| |\bm k|^3 \frac{(9-20 a_0^2 |\bm k|^2)^2 }{(4 a_0^2 |\bm k|^2+9)^8}
		\begin{pmatrix}
		(1-a^2)e^{-\frac{1}{2}\sigma^2( |\bm k|-\Omega)^2} & a \sqrt{1-a^2} e^{-\frac{1}{2}\sigma^2( |\bm k|^2 + \Omega^2)} \\
		a \sqrt{1-a^2} e^{-\frac{1}{2}\sigma^2( |\bm k|^2 + \Omega^2)} & a^2 e^{-\frac{1}{2}\sigma^2( |\bm k|+\Omega)^2} \\
		\end{pmatrix},
		\end{aligned}\\
		&\left.\tr_{\textsc{f}}\left(\hat  U^{(2)} \hat \rho_i\right)\right|_{\openone}\nonumber \\
		&\begin{aligned}
		= -\frac{331776}{\pi} (e a_0 \sigma)^2 \int_0^{\infty} \dd|\bm k| |\bm k|^3 \frac{16 a_0^4 |\bm k|^4-8 a_0^2 |\bm k|^2+9 }{(4 a_0^2 |\bm k|^2+9)^8} &\left[\text{erfc}\left(\frac{\ii \sigma(|\bm k|+\Omega)}{\sqrt{2}}\right) e^{-\frac{1}{2} \sigma^2(|\bm k|+\Omega)^2} \begin{pmatrix}
		a^2 & a \sqrt{1-a^2} \\
		0 & 0 \\
		\end{pmatrix} \right. \\
		&~\left. +\text{erfc}\left(\frac{\ii \sigma(|\bm k|-\Omega)}{\sqrt{2}}\right) e^{-\frac{1}{2} \sigma^2(|\bm k|-\Omega)^2} \begin{pmatrix}
		0 & 0 \\
		a \sqrt{1-a^2} & 1-a^2 \\
		\end{pmatrix} \right], 
		\end{aligned}\\
		&\left.\tr_{\textsc{f}}\left(\hat  U^{(2)} \hat \rho_i\right)\right|_{\bm k \otimes \bm k}\nonumber\\
		&\begin{aligned}
		= \frac{12288}{\pi} (e a_0 \sigma)^2 \int_0^{\infty} \dd|\bm k| |\bm k|^3 \frac{(9-20 a_0^2 |\bm k|^2)^2 }{(4 a_0^2 |\bm k|^2+9)^8}& \left[\text{erfc}\left(\frac{\ii \sigma(|\bm k|+\Omega)}{\sqrt{2}}\right) e^{-\frac{1}{2} \sigma^2(|\bm k|+\Omega)^2} \begin{pmatrix}
		a^2 & a \sqrt{1-a^2} \\
		0 & 0 \\
		\end{pmatrix} \right. 
		\\
		&~\left.  + \text{erfc}\left(\frac{\ii \sigma(|\bm k|-\Omega)}{\sqrt{2}}\right) e^{-\frac{1}{2} \sigma^2(|\bm k|-\Omega)^2} \begin{pmatrix}
		0 & 0 \\
		a \sqrt{1-a^2} & 1-a^2 \\
		\end{pmatrix} \right],
		\end{aligned}
		\end{align}
		where we have used the following for the nested time integrals:
		\begin{align}
		&\int_{-\infty}^{\infty} \dd t \int_{-\infty}^{t} \dd t' e^{\pm \ii \Omega (t-t')} e^{-\frac{t^2}{\sigma^2}} e^{-\frac{t'^2}{\sigma^2}} e^{-\ii |\bm k| (t-t')}= \sqrt{\pi}\frac{\sigma}{2} \int_{-\infty}^{\infty} \dd t e^{-\frac{1}{4} (|\bm k| \mp \Omega)^2 \sigma^2} e^{\ii (\pm \Omega-|\bm k|) t} e^{-\frac{t^2}{\sigma^2}} \nonumber \\
		&= \frac{\pi \sigma^2}{2} e^{-\frac{1}{2}(\Omega^2 +  |\bm k|^2) \sigma^2} e^{\pm |\bm k| \Omega \sigma^2} I\left(\frac{\sigma}{2}(|\bm k|\mp\Omega),\sigma(|\bm k|\mp \Omega) \right) = \frac{\pi \sigma^2}{2} \text{erfc}\left(\frac{\ii \sigma (|\bm k| \mp \Omega)}{\sqrt{2}}\right) e^{\frac{1}{2}(-|\bm k|^2 \sigma^2 \pm |\bm k| \Omega \sigma^2- \Omega^2 \sigma^2)} \nonumber \\
		&=\frac{\pi \sigma^2}{2} \text{erfc}\left(\frac{\ii \sigma (|\bm k| \mp \Omega)}{\sqrt{2}}\right) e^{\frac{1}{2}\sigma^2 (|\bm k| \mp \Omega)^2},
		\end{align}
		with 
		\begin{align}
		I(a,b)= \int_{-\infty}^{\infty} \dd x e^{-a^2-\ii b x - x^2} \text{erf}(x-\ii a)=-\ii \sqrt{\pi} e^{-a^2-\frac{b^2}{4}} \text{erf}\left(\frac{a+\frac{b}{2}}{\sqrt{2}} \right).
		\end{align}
		Then the whole change in the atomic state is 
		\begin{align}
		\Delta \hat \rho^\textsc{g}=&\frac{ 24576 (a_0 e \sigma)^2}{ \pi}\int_0^{\infty} \dd |\bm k| \frac{|\bm k|^3 e^{-\frac{1}{2} \sigma^2 (|\bm k|+ \Omega)^2} }{\left(4 a_0^2 |\bm k|^2 +9\right)^6} \left\{ 2 \left[(1-a^2) e^{2 |\bm k| \sigma^2 \Omega}-a^2 \right] \begin{pmatrix}
		1&0 \\
		0 & -1  \\
		\end{pmatrix} \right. \nonumber \\
		&\quad+\left. \left( a \sqrt{1-a^2} \left[e^{2 |\bm k| \sigma^2 \Omega} \text{erf}\left(\frac{\ii \sigma (|\bm k| - \Omega)}{\sqrt{2}} \right)- \text{erf}\left(\frac{\ii \sigma (|\bm k| + \Omega)}{\sqrt{2}} \right) -\left(1- e^{|\bm k| \sigma^2 \Omega}\right)^2 \right] \begin{pmatrix}
		0&0 \\
		1 & 0  \\
		\end{pmatrix}
		+ \text{H.c.} \right) \right\}.
		\end{align}

		For the sudden top-hat switching $(ii)~ \chi^\textsc{s}(t)=\Theta(t)\Theta(-t+\sigma)$, with the duration of interaction $\sigma$,
		\begin{align}
		\left. \tr_{\textsc{f}}\left( \hat U^{(1)} \hat \rho_i \hat U^{(1) \dagger}\right)\right|_{\openone}=& \frac{1327104}{\pi^2} (e a_0)^2 \int_0^{\infty} \dd|\bm k| |\bm k|^3  \frac{16 a_0^4 |\bm k|^4-8 a_0^2 |\bm k|^2+9 }{(4 a_0^2 |\bm k|^2+9)^8}  \left[  \frac{ 1-\cos{\left(\sigma( |\bm k|-\Omega) \right)}}{(|\bm k|-\Omega)^2} \begin{pmatrix}
		1-a^2 & 0 \\
		0 & 0 \\
		\end{pmatrix}\right. \nonumber
		\\
		& \quad\left.  + \frac{ 1-\cos{\left(\sigma( |\bm k|+\Omega) \right)}}{(|\bm k|+\Omega)^2} \begin{pmatrix}
		0 & 0 \\
		0 & a^2 \\
		\end{pmatrix} 
		- a \sqrt{1-a^2} \frac{\cos{(\sigma |\bm k|)}-\cos{(\sigma \Omega)}}{|\bm k|^2+\Omega^2} \begin{pmatrix}
		0 & e^{-\ii \Omega \sigma} \\
		e^{\ii \Omega \sigma} & 0 \\
		\end{pmatrix} \right],\\
		\left.\tr_{\textsc{f}}\left( \hat U^{(1)} \hat \rho_i \hat U^{(1) \dagger}\right)\right|_{\bm k \otimes \bm k}=& -\frac{49152}{\pi^2} (e a_0)^2 \int_0^{\infty} \dd|\bm k| |\bm k|^3 \frac{(9-20 a_0^2 |\bm k|^2)^2 }{(4 a_0^2 |\bm k|^2+9)^8}  \left[\frac{ 1-\cos{\left(\sigma( |\bm k|-\Omega) \right)}}{(|\bm k|-\Omega)^2} \begin{pmatrix}
		1-a^2 & 0 \\
		0 & 0 \\
		\end{pmatrix}\right. \nonumber
		\\
		&\quad \left. + \frac{ 1-\cos{\left(\sigma( |\bm k|+\Omega) \right)}}{(|\bm k|+\Omega)^2} \begin{pmatrix}
		0 & 0 \\
		0 & a^2 \\
		\end{pmatrix} 
		- a \sqrt{1-a^2} \frac{\cos{(\sigma |\bm k|)}-\cos{(\sigma \Omega)}}{|\bm k|^2+\Omega^2} \begin{pmatrix}
		0 & e^{-\ii \Omega \sigma} \\
		e^{\ii \Omega \sigma} & 0 \\
		\end{pmatrix} \right], \\
		\left.\tr_{\textsc{f}}\left(\hat  U^{(2)} \hat \rho_i\right)\right|_{\openone}=& \frac{663552}{\pi^2} (e a_0)^2 \int_0^{\infty} \dd|\bm k| |\bm k|^3 \left[\frac{\ii \sigma(|\bm k|+\Omega)+ e^{-\ii \sigma(|\bm k|+\Omega)}-1}{(|\bm k|+\Omega)^2} \begin{pmatrix}
		a^2 & a \sqrt{1-a^2} \\
		0 & 0 \\
		\end{pmatrix}\nonumber \right.
		\\
		&\quad + \left.\frac{\ii \sigma(|\bm k|-\Omega)+ e^{-\ii \sigma(|\bm k|-\Omega)}-1}{(|\bm k|-\Omega)^2} \begin{pmatrix}
		0 & 0 \\
		a \sqrt{1-a^2} & 1-a^2 \\
		\end{pmatrix} \right]  \frac{16 a_0^4 |\bm k|^4-8 a_0^2 |\bm k|^2+9 }{(4 a_0^2 |\bm k|^2+9)^8},\\
		\left.\tr_{\textsc{f}}\left(\hat  U^{(2)} \hat \rho_i\right)\right|_{\bm k \otimes \bm k}=& -\frac{24576}{\pi^2} (e a_0)^2 \int_0^{\infty} \dd|\bm k| |\bm k|^3 \frac{(9-20 a_0^2 |\bm k|^2)^2 }{(4 a_0^2 |\bm k|^2+9)^8}  \left[\frac{\ii \sigma(|\bm k|+\Omega)+ e^{-\ii \sigma(|\bm k|+\Omega)}-1}{(|\bm k|+\Omega)^2} \begin{pmatrix}
		a^2 & a \sqrt{1-a^2} \\
		0 & 0 \\
		\end{pmatrix}\nonumber \right. 
		\\
		&\quad\left.  + \frac{\ii \sigma(|\bm k|-\Omega)+ e^{-\ii \sigma(|\bm k|-\Omega)}-1}{(|\bm k|-\Omega)^2} \begin{pmatrix}
		0 & 0 \\
		a \sqrt{1-a^2} & 1-a^2 \\
		\end{pmatrix} \right].
		\end{align}
		Hence we find the total change of the atomic state to be
		\begin{align}
		\Delta \hat \rho^\textsc{s}=& \frac{49152 (a_0 e)^2}{\pi^2} \int_0^{\infty} \!\dd |\bm k| \frac{ |\bm k|^3 }{\left( 4 a_0^2 |\bm k|^2 +9 \right)^6 (|\bm k|^2-\Omega^2)^2} \left\{\vphantom{\begin{pmatrix}
			1 & 0 \\
			0 & -1 \\
			\end{pmatrix}} \left(2a^2 (|\bm k|-\Omega )^2 \cos  (\sigma  (|\bm k|+\Omega )) \right.\right. \nonumber \\
		&\quad\quad \left.\left. +2\left(1-2 a^2\right) \left(|\bm k|^2+\Omega^2\right)+4 |\bm k| \Omega+2(a^2-1\right)  \cos  (\sigma  (|\bm k|-\Omega ))(|\bm k|+\Omega )^2 \right)  \begin{pmatrix}
		1 & 0 \\
		0 & -1 \\
		\end{pmatrix}  \nonumber \\
		&\,\quad+ \left( \vphantom{\begin{pmatrix}
			1 & 0 \\
			0 & -1 \\
			\end{pmatrix}} \left[ e^{2 \ii \sigma \Omega}(|\bm k|^2 -\Omega^2) +|\bm k|^2 (2 \ii \sigma \Omega-1) +  4 \Omega e^{\ii \sigma \Omega} ( \Omega \cos{(|\bm k| \sigma)} - \ii |\bm k| \sin{(|\bm k| \sigma)}) - \Omega^2 (3 + 2\ii \sigma \Omega)\right] \right. \nonumber \\
		&\left. \left.\quad \quad  \times a \sqrt{1-a^2}  \begin{pmatrix}
		0 & 0 \\
		1 & 0 \\
		\end{pmatrix}  +  \text{H.c.} \right)\right\}.
		\end{align}
		
		We can particularize to the gapless case ($\Omega=0$), allowing then to perform the last remaining integral such that
		\begin{align}
		\Delta \hat \rho^\textsc{s}_{\Omega=0}=& \frac{512 e^2 }{295245 \pi^2}\left(1- 2 a^2 \right) \left[24-\sqrt{\pi} \MeijerG*{2}{1}{1}{3}{0}{0,5,\frac{1}{2}}{\frac{9 \sigma^2}{16 a_0^2}}\right]  \begin{pmatrix}
		1 & 0 \\
		0 & -1 \\
		\end{pmatrix} ,
		\end{align}
		where $\MeijerG*{m}{n}{p}{q}{a_1, \dots, a_p}{b_1, \dots, b_q}{z}$ is the Meijer G-function.

		Finally for $(iii)~ \chi^\textsc{d}(t)=C \delta(t)$, where $C$ is some constant  with mass dimension  $\left[C \right]=-1$,
		\begin{align}
		\left. \tr_{\textsc{f}}\left( \hat U^{(1)} \hat \rho_i \hat U^{(1) \dagger}\right)\right|_{\openone}=& \frac{663552 C^2}{\pi^2} (e a_0)^2 \int_0^{\infty} \dd|\bm k| |\bm k|^3 \frac{16 a_0^4 |\bm k|^4-8 a_0^2 |\bm k|^2+9 }{(4 a_0^2 |\bm k|^2+9)^8}  \begin{pmatrix}
		1-a^2 & a \sqrt{1-a^2} \\
		a \sqrt{1-a^2} & a^2 \\
		\end{pmatrix}, \\
		\left.\tr_{\textsc{f}}\left( \hat U^{(1)} \hat \rho_i \hat U^{(1) \dagger}\right)\right|_{\bm k \otimes \bm k}=& -\frac{24576 C^2}{\pi^2} (e a_0)^2 \int_0^{\infty} \dd|\bm k| |\bm k|^3 \frac{(9-20 a_0^2 |\bm k|^2)^2 }{(4 a_0^2 |\bm k|^2+9)^8}  \begin{pmatrix}
		1-a^2 & a \sqrt{1-a^2} \\
		a \sqrt{1-a^2} & a^2 \\
		\end{pmatrix}, \\
		\left.\tr_{\textsc{f}}\left(\hat  U^{(2)} \hat \rho_i\right)\right|_{\openone}=&- \frac{331776 C^2}{\pi^2} (e a_0)^2 \int_0^{\infty} \dd|\bm k| |\bm k|^3 \frac{16 a_0^4 |\bm k|^4-8 a_0^2 |\bm k|^2+9 }{(4 a_0^2 |\bm k|^2+9)^8}  \begin{pmatrix}
		a^2 & a \sqrt{1-a^2} \\
		a \sqrt{1-a^2} & 1-a^2 \\
		\end{pmatrix},\\
		\left.\tr_{\textsc{f}}\left(\hat  U^{(2)} \hat \rho_i\right)\right|_{\bm k \otimes \bm k}=& \frac{12288 C^2}{\pi^2} (e a_0)^2 \int_0^{\infty} \dd|\bm k| |\bm k|^3 \frac{(9-20 a_0^2 |\bm k|^2)^2 }{(4 a_0^2 |\bm k|^2+9)^8} \begin{pmatrix}
		a^2 & a \sqrt{1-a^2} \\
		a \sqrt{1-a^2} & 1-a^2 \\
		\end{pmatrix}.
		\end{align}
		The nested time integrals over the two Dirac distributions are mathematically ambiguous (see \cite{Pozas2017}) and require us to understand them as some sort of limit of a sequence of functions. If the delta distribution is understood as the limit of a sequence of symmetric peaked functions of smaller and smaller width and constant area (e.g., the Dirac distribution is the short width limit ---symmetrically taken--- of a sudden top-hat or Gaussian function), as it is shown in the appendix of  \cite{Pozas2017}, one can show that
		\begin{align}
		\int_{-\infty}^{\infty} \dd t \int_{-\infty}^t \dd t' e^{- \ii |\bm k| (t-t')} e^{\pm \ii \Omega (t-t')} \delta(t) \delta(t')= \frac{1}{2}.
		\end{align}
		Our result can be integrated analytically and hence the change in the atomic state due to the delta switching is
		\begin{align}
		\Delta \hat \rho^\textsc{d}= \frac{128 C^2 e^2}{10935 \pi^2 a_0^2} \left(1-2 a^2\right) \begin{pmatrix}
		1&0 \\
		0 & -1  \\
		\end{pmatrix}.
		\end{align}
		In an equal superposition ($a=\frac{1}{\sqrt{2}}$) the atomic state does not get perturbed to second order in perturbation theory for the delta and gapless sudden switching. Thus the purity is preserved which yields $H_{\text{min}}=1~\text{bit}$.
		As can be easily checked, for all cases the perturbation is traceless and Hermitian.

	\subsubsection{Assessing the validity of the perturbative approach}\label{numvalues}
	
	In the following we will present numerical results for the change in the atomic state after interaction for the different switching functions. We will focus on regimes of low min-entropy, establishing that perturbation theory indeed holds in these cases below a certain threshold of the coupling strength $e$ depending on the switching function. Since the excited state always led to the least generated  $H_{\text{min}}$, we will particularize to $a=0$ that will yield the worst-case-scenario for perturbation theory to hold. Assuming this, the initial state reads
	\begin{align}
	    \hat{\rho}_0=\left(
\begin{array}{cc}
0 & 0 \\
 0 & 1 \\
\end{array}
\right).
	\end{align}
	We saw earlier (Fig.~\ref{gausspara}a, \ref{unit}b, and \ref{dirac1}a) that increasing the coupling strength can drastically reduce the generated randomness. We choose in the following $a_0=2.68\cdot 10^{-4}~\text{eV}^{-1}$, $\Omega =3.73~\text{eV}$, and $\sigma=2.5 \times 10^{-3}$. Then we find that the coupling strength needs to be below the following values such that the magnitude of the perturbation to the initial state is at most $0.1$ times the original state for Gaussian and gapless sudden switching:
	\begin{align}
	    \Delta \hat \rho^\textsc{g}(e=12.8)&= \left(
\begin{array}{cc}
 0.1 & 0 \\
 0 & -0.1 \\
\end{array}
\right),\\
	    \Delta \hat \rho^\textsc{h}_{\Omega=0}(e=4.7)&=\left(
\begin{array}{cc}
 0.1 & 0 \\
 0 & -0.1 \\
\end{array}
\right),  
\end{align}
with both having a purity of $\tr\left(\hat \rho_{\textsc{a}}^2\right)=0.82$. We declare that the end-limit of the safe applicability of the perturbative analysis.

On the other end, for the largest perturbation shown in the plots in the case of Dirac switching (Fig.~\ref{dirac1}a)
\begin{align}
	    \Delta \hat \rho^\textsc{d}(e=4 \times 8.54 \times 10^{-2})&= \left(
\begin{array}{cc}
 0.012 & 0. \\
 0. & -0.012 \\
\end{array}
\right), ~~\tr\left(\hat \rho_{\textsc{a}}^2\right)=0.98,
	\end{align}
	so we stay well within the regime of perturbation theory for all regimes studied of delta switching.

However, for Gaussian and sudden switching in the plots Fig.~\ref{gausspara}a and Fig.~\ref{unit}b we have gone slightly above those `safe' numbers. The changes in the density matrices in the worst-case scenario for the plotted figures are
\begin{align}
	    \Delta \hat \rho^\textsc{g}(e=17)&= \left(
\begin{array}{cc}
 0.177 & 0. \\
 0. & -0.177 \\
\end{array}
\right), \\
	    \Delta \hat \rho^\textsc{h}_{\Omega=0}(e=7)&=\left(
\begin{array}{cc}
 0.22 & 0 \\
 0 & -0.22 \\
\end{array}
\right).
\end{align}
Therefore they are still under control, even if they are outside of the perturbative regime (higher order corrections will still be smaller).
	
		\subsubsection{Symmetry property of the min-entropy}\label{symm}
		In order to see why to leading order in perturbation theory $H_{\text{min}}$ is symmetric around $a=\frac{1}{\sqrt{2}}$ for Dirac and gapless ($\Omega=0$) sudden switching, let us focus on the term
		\begin{align}
		    \left(1-2 a^2\right) \begin{pmatrix}
		1&0 \\
		0 & -1  \\
		\end{pmatrix},\label{transformation}
		\end{align}
		which is present in \eqref{gapless} and \eqref{diracfinal}. Under the transformation $a^2 \rightarrow 1-a^2$, \eqref{transformation} yields an additional factor of $(-1)$. Since the min-entropy (see \eqref{hmin}) depends on the purity of the state after the interaction,
		\begin{align}
		    		\tr\left(\hat \rho_\textsc{a}^2\right)=\tr\left((\hat \rho_{\textsc{a},i}+\Delta \hat \rho)^2\right)=\tr\left(\hat \rho_{\textsc{a},i}^2 + 2 \hat \rho_{\textsc{a},i} \Delta \hat \rho \right) + \mathcal{O}(e^4), 
		\end{align}
		the object of interest is, using \eqref{initi}, 
		\begin{align}
		    \tr\left(\hat \rho_{\textsc{a},i} \Delta \hat \rho\right)\propto  \left[a^2 -(1-a^2)\right] (1-2 a^2)=- (1-2 a^2)^2. \label{neu}
		\end{align}
		Clearly \eqref{neu} is then invariant under $a^2 \rightarrow 1-a^2$. Therefore, we expect the min-entropy to be symmetric around $a=\frac{1}{\sqrt{2}}$ in the case for gapless sudden and Dirac switching, as is shown in Fig.~\ref{unit}a and \ref{dirac1}a.

		On the other hand, for Gaussian switching there is none such invariance of the min-entropy.
		Starting from eq.~\eqref{gaussian}, one sees that the off-diagonal elements are invariant under $a^2 \rightarrow 1-a^2$, the diagonal elements however would yield an additional factor of $(-1)$ (as needed for symmetry of the min-entropy under the transformation) if it were not for the factor $\exp(2 |\bm k| \sigma^2 \Omega)$. It is therefore the case that if we use a non-homogeneous switching function (there is real time variation of the coupling) for atoms with non-zero internal dynamics ($\Omega>0$) the free time evolution distorts the symmetry of the plots.


		\end{widetext}
	\bibliography{myref}

\end{document}